\documentclass[12pt,a4paper]{article}
\usepackage[a4paper,top=3cm,bottom=3cm,left=2cm,right=2cm,bindingoffset=5mm]{geometry}
\usepackage[usenames, dvipsnames]{color}
\usepackage[greek,english]{babel}
\usepackage{amsmath}
\usepackage{amssymb}
\usepackage[usenames,dvipsnames]{color}
\usepackage{cancel}
\usepackage{graphicx}
\usepackage{mathrsfs}
\usepackage{bbm}
\usepackage{cite}
\usepackage{hyperref}

\hypersetup{
colorlinks=true,         
linkcolor=blue,          
citecolor=red,        
urlcolor=red            
}

\def\sfrac{\textstyle \frac}
\def\d{\mathrm{d}}

\begin{document}

\title{\vspace{-2.5cm}\bf Through the Big Bang\\in inflationary cosmology}

\author{Flavio Mercati\thanks{flavio.mercati@gmail.com}\\
Dipartimento di Fisica ``Ettore Pancini'',\\
Universit\`{a} di Napoli {\sl Federico~II}, Napoli, Italy\\
and
INFN, Sezione di Napoli, Italy}

\date{}

\maketitle

\vspace{-1cm}

\begin{abstract}
Singularities in General Relativity are regions where the description of spacetime in terms of a pseudo-Riemannian geometry breaks down. The theory seems unable to predict the evolution of the physical degrees of freedom around and beyond such regions. In a recent paper, the author and collaborators challenged this view by providing an example of a singularity at which Einstein's equations can be rewritten in a form that satisfies an existence and uniqueness theorem, thereby predicting that each solution can be continued uniquely through the singularity. This result was obtained under the assumption of homogeneity (but not isotropy), and requires the presence of a massless free scalar field. This paper extends the result to N scalar fields with a potential, the only requirement being that it does not grow too fast. In particular, the result is compatible with inflationary potentials, \emph{e.g.} Starobinsky's. This brings us one step closer to the goal of extending the original result to realistic cosmologies.
\end{abstract}

\tableofcontents

\section{Introduction}

The evolution of regular initial conditions into singularities is one of the most striking implications of General Relativity. The status of singularities (whether they are physical or not) was subject to much debate, but after Penrose and Hawking formulated their singularity theorems~\cite{PenroseSingularityTheorem,HawkingSingularityTheorem1,HawkingSingularityTheorem2,HawkingSingularityTheorem3,
HawkingPenroseSingularityTheorem,HawkingEllisBook} a consensus seems to have formed.\footnote{Before~\cite{PenroseSingularityTheorem,HawkingSingularityTheorem1,HawkingSingularityTheorem2,HawkingSingularityTheorem3,
HawkingPenroseSingularityTheorem,HawkingEllisBook}, people speculated that singularities only form in exactly symmetric solutions. For example, it was suggested that, in the gravitational collapse of matter, only zero-angular momentum solutions would be able to form a singularity, matching what happens in Newtonian gravity, where a total collision is forbidden by a nonzero angular momentum.
 The singularity theorems  prove such suggestions wrong.} Singularities are, at least in classical General Relativity, regarded as a physical prediction of the theory, which moreover is unavoidable in the case of expanding/contracting initial conditions with matter satisfying reasonable energy conditions. The most popular proposals for a resolution of general-relativistic singularities nowadays rely on quantum effects~\cite{MisnerQuantumCosmologyI-1969,Berger1982,AshtekarBojowald}, taking inspiration from the resolution, in quantum electrodynamics,  of the infinite self-energy problem of point-charges~\cite{Schwinger1948}.

The Penrose--Hawking singularity theorems assume, as notion of singularity, some form of geodesic incompleteness~\cite{HawkingEllisBook}. This tells us that the spacetime manifold has singular points  which can be reached in finite proper time and at which some components of the spacetime curvature become infinite in a coordinate-independent way. This typically implies that an observer following a geodesic would experience infinite tidal stresses. This is often considered a situation in which the theory loses predictivity because there is no prescription to continue geodesics past that point. Since the classical equations of motion of known fields `break down' there, one can, in Hawking's words, put it more strikingly: ``One does not know what will come out of a singularity''~\cite{HawkingPRD1976}.

In a recent paper~\cite{ThroughTheBigBang} the author and collaborators challenged the idea that gravitational singularities generally mark a loss of predictivity of the classical theory. In a class of homogeneous anisotropic cosmological models we showed it is possible to find a set of variables in which Einstein's equations (which reduce to ordinary differential equations (ODEs) under the homogeneous ansatz) satisfy the existence and uniqueness theorem of ODEs at the Big Bang singularity. The class of models for which this property was proven is Bianchi IX universes (spatially compact with $S^3$ topology) with \emph{stiff} matter sources. The requirement of  a stiff equation of state, $p = \epsilon$, is realized for example in the case of a free (massless and zero-potential, but of course minimally-coupled) scalar field. In the absence of stiff matter the Bianchi IX system evolves according to Misner's `mixmaster' dynamics~\cite{MTW}, which is qualitatively equivalent to that of a billiard ball in a triangular pool table. In Misner coordinates, the singularity happens in the infinite past (or future), and the billiard-ball bounces never stop. However, the system takes a finite proper time to reach the singularity, which means that the bounces happen at an accelerated pace (in proper time) as the singularity is approached, and infinitely many bounces take place in a finite amount of proper time~\cite{MTW}. Therefore, the dynamical variables (which are the Misner variables measuring the anisotropy of spatial slices)  do not admit a well-defined limit at the singularity, just as $\lim_{x \to 0} \sin (1/x)$ does not converge. To avoid this, it is necessary to introduce stiff matter. Normally the back-reaction of matter sources on the geometric degrees of freedom is negligible near the singularity (``matter doesn't matter at the Big Bang''~\cite{Matterdoesntmatter}), and this is in particular true of Standard Model matter. But in the case of stiff matter like a massless scalar field, its contribution to the dynamics can never be neglected, and in particular it implies that the system, in its evolution towards the singularity, will necessarily enter a phase of `quiescence'~\cite{Barrow1978}. This means that after a finite number of billiard-ball bounces the system stabilizes around a free evolution without any further bounce all the way to the singularity. This is a necessary condition for being able to continue the evolution past the singularity: the geometrical degrees of freedom must admit well-defined limits -- in this case, they asymptote to a degenerate, effectively one-dimensional geometry, in which two of the spatial directions are infinitely squeezed with respect to the other.\footnote{This is the generic situation, but there is also a measure-zero set of solutions in which one direction gets infinitely squeezed while the other two keep a finite ratio, so the geometry becomes effectively two-dimensional.}  These degenerate geometries cannot support a nonzero volume, and this is why the singularity takes place there. As a matter of fact, if we exclude the exceptional case of isotropic FRLW solutions whose spatial geometry remains fixed apart from a change of scale, all solutions of the `quiescent Bianchi IX' system can only reach the singularity at one of the degenerate geometries.

As I mentioned, the system, written in terms of the appropriate variables, now satisfies  the Picard--Lindel\"of theorem~\cite{teschl2012ordinary} on the existence and uniqueness of solutions to ODEs everywhere, and in particular at the singularity. The most striking implication of this is that the orientation of spatial slices will have to change at the singularity. This is a consequence of the fact that, in order to write the equations of motion in a way that satisfies the existence and uniqueness theorem, one has to assume a definite global topology for configuration space and this places oppositely-oriented spatial geometries next to each other, separated by the degenerate geometries, which act like a boundary between opposite orientations and, as I remarked above, are the only place where the volume can vanish and the singularity can take place. In this way each solution looks like two expanding universes with opposite orientations glued at the Big Bang, where the orientation flip takes place. As the author and collaborators argued in~\cite{ArrowPaper}, there is a well-defined sense in which we can talk about a time-reversal at the Big Bang in such a situation. In fact, if unbiased initial conditions are chosen at the Big Bang, the familiar arrows of time (in particular the thermodynamic one) will be seen to emerge, as the universe evolves away from the Big Bang, in the time direction in which the universe is expanding. This argument supports the idea that the overwhelming majority of the solutions of our system are PT-invariant, in the sense that the two halves of each solution have opposite spatial orientation (P) and oppositely-pointing arrows of time (T). It is natural, at this point, to speculate that they will be fully CPT-symmetric, but in order to check this out explicitly we would need to introduce charged matter sources and discuss their behaviour at the singularity.

On a more conceptual level, the result of~\cite{ThroughTheBigBang} shows that solutions can be continued uniquely and \emph{predictively} through the Big Bang. It seems that, at least  for this type of singularity and under the homogeneous ansatz we made, one \emph{does} in fact know what will come out of a singularity. The question remains how general this result is. The main limitation of~\cite{ThroughTheBigBang}  is of course the assumption of homogeneity. We have, however, encouraging indications coming from work in mathematical relativity on the Belinski--Kalatnikov--Lifshitz conjecture~\cite{BKL}. This conjecture suggests that the dynamics of General Relativity near a singularity reduces to that of an infinite collection of Bianchi IX models, decoupled from each other, one per spatial point (or, alternatively, one per Fourier mode of the metric field). This is an unproven conjecture, which however has gained, over the years, a significant amount of numerical support~\cite{Garfinkle2004,Curtis2005}. The closest we got to a proof of the conjecture is, in my opinion, the result by Andersson and Rendall~\cite{AR}, which, strikingly, relates to \emph{quiescent} solutions of Einstein's equations. In the quiescent/stiff matter case, the authors are able to prove the conjecture for analytic solutions of Einstein's equations. This is a strong indication that~\cite{ThroughTheBigBang} can be generalized to inhomogeneous solutions, essentially by promoting our dynamical variables to functions of the spatial point. In the general case Einstein's equations will be partial differential equations which couple those variables at nearby points, but as the system approaches the singularity and quiescence sets in the coupling between different points will be suppressed, and the system will tend to a set of ultralocal (systems of) ODEs, one per spatial point. The continuation result will then hold in a pointwise fashion. Explicit investigation of this conjecture will be left to future studies.

Another aspect that requires further investigation is the matter content of the model. In~\cite{ThroughTheBigBang} only a massless scalar was considered. In fact any other type of Standard Model matter is expected to decouple before the singularity is approached (decouple in the sense that the geometrical degrees of freedom stop being sourced by the matter ones, while the opposite continues to hold -- matter `piggybacks' on geometry)~\cite{Matterdoesntmatter}. This is something that can be verified for perturbations around FRLW solutions, but is not proven in general nonperturbative solutions that are far from isotropy, and it is therefore a further direction for future investigation. Under the homogeneous  ansatz and with an $S^3$ topology, however, only scalar fields can have a nonzero value, this being so for topological reasons (there are no constant nonzero vector or spinor fields on a 3-sphere). So the investigation of gauge or fermion fields necessarily goes first through the generalization to inhomogeneous situations.

The natural next step is, then, to retain the homogeneous ansatz but to relax the assumption of a free, zero-mass and zero-potential scalar field by introducing a generic potential and studying under what conditions quiescence can be attained and, in such cases, whether the continuation through the singularity result still holds. This will be the subject of the present paper. The main phenomenological interest in this generalization comes from the possibility that the scalar field that is responsible for the onset of quiescence might be the inflaton. I will show, in the following, that such a scenario is possible and does not require any restrictive assumption on the form of the inflationary potential or on the initial conditions of the solutions. In other words, generic solutions of inflationary models exhibit quiescence and can be continued uniquely through the Big Bang singularity. As an explicit example I will consider Starobinsky's model, which is one of the most popular inflationary models and is in full accordance with the present experimental constraints.

In addition to this, I will also introduce a simpler set of variables that accomplishes the main result of~\cite{ThroughTheBigBang}, \emph{i.e.} showing that the existence and uniqueness theorem holds at the singularity, but in a more intuitive way, which in turn is also amenable to generalizations to models with an arbitrary number of fields. I will show this by using the new variables to generalize the result to $N$ scalar fields.

\section{The model}

I will be working within the framework of the Arnowitt--Deser--Misner Hamiltonian formulation of GR~\cite{MTW,FlavioSDbook}. The dynamical variables are then the Euclidean-signature 3D metric $g_{ij}$ induced on a spacelike hypersurface and its conjugate momentum $p^{ij}$, which is related to the extrinsic curvature of  the hypersurface $K^{ij}$ by a Legendre transform $p^{ij} = \sqrt g \left( \text{tr} K - K^{ij} \right)$. If we introduce also a scalar field, its dynamical variables will simply be the field $\phi$ and its conjugate momentum $\pi_\phi$. The symplectic potential is then
\begin{equation}\label{ADM-symplectic-potential}
\Theta = \int \d^3x \left( p^{ij} \delta g_{ij} + \pi_\phi \delta \phi \right) \,.
\end{equation}
The theory is constrained: its Hamiltonian vanishes on-shell because it is a linear combination [through Lagrange multipliers which are four components of the spacetime metric whose time derivatives does not appear in the Einstein--Hilbert action, the \emph{lapse} $N(x)$ and the \emph{shift} $N_i(x)$] of the following functional constraints:
\begin{equation}\label{ADM-constraints}
\begin{aligned}
\mathcal H [N]  &=  \int \d^3 x N \left[ \sqrt g \, R  - {\sfrac 1 {\sqrt g}} \left( p^{ij} p_{ij} - {\sfrac 1 2} (\text{tr} p)^2 +  {\sfrac 1 2}  \pi_\phi^2 \right) - {\sfrac 1 2} \sqrt g \, g_{ij} \nabla^i \phi \nabla^j \phi - \sqrt g \, V(\phi) \right]\,,
\\
\mathcal D[N_i] &= \int \d^3 x N_i \left[- 2 \nabla_j p^{ij} + \pi_\phi \nabla^i \phi \right]  \,.
\end{aligned}
\end{equation}

My initial assumption will be that there exists a hypersurface of simultaneity on which the spatial metric, as well as all the other fields, is homogeneous, and that the topology of this hypersurface is $S^3$. Then we can write the spatial line element as
\begin{equation}\label{3DLineElement}
ds^2 = \delta_{ab} \, e^a \, e^b = \delta_{ab}  e^a_i \, e^b_j \, dx^i dx^j \,,
\end{equation}
where $e^a_i$ are frame fields (triads), which can be written as (see Appendix~\ref{Appendix_A} for the explicit derivation)
\begin{equation}\label{Translation-invariant-triads}
e^1 = \pm v^{1/3} \, e^{- \frac{x}{2\sqrt{2}} + \frac{y}{2\sqrt{6}} }  \sigma^1 \,,  ~~~
e^2 = \pm v^{1/3} \,  e^{ \frac{x}{2\sqrt{2}} + \frac{y}{2\sqrt{6}} }  \sigma^2 \,, ~~~
e^3 = \pm v^{1/3} \,  e^{ - \frac{y}{2\sqrt{6}} } \sigma^3 \,,
\end{equation}
where now $\sigma^a$ are the three translationally-invariant one-forms on $S^3$:
\begin{equation}\label{Translation-invariant-one-forms-S3}
\left\{ \begin{aligned}
\sigma^1 &=  \sin r \, \d \theta - \cos r \, \sin \theta \, \d \varphi
\\
\sigma^2 &=  \cos r \, \d \theta + \sin r \, \sin \theta \, \d \varphi 
\\
\sigma^3 &=  - \d r   - \cos   \theta \, \d \varphi 
\end{aligned} \right.  \,,
\end{equation}
where I introduced the local coordinates on $S^3$ $ r \in (0,\pi)$, $\theta \in (0,\pi)$, $\varphi \in (0,2\pi]$. (\ref{Translation-invariant-triads}) is the most general translation-invariant triad on $S^3$, and it depends on three real variables $x$, $y$ and $v$. The first two, called Misner variables, are \emph{anisotropy parameters}. When they are both zero the metric is homogeneous and isotropic, and represents the round metric on the three-sphere. Any other value of $x$ and $y$ describes a metric which, albeit homogeneous, has preferred directions. For example, two great circles (geodesics)  starting from the same point but in different directions can have different lengths. The variable $v = |\det e|/\sin \theta$ represents the volume of our spatial slices, while  $x$ and $y$ determine the \emph{conformal geometry} of the slices. Their Riemannian geometry is specified once we have the value of $v$ too.
Notice that the metric $g_{ij}$, being quadratic in $e^a_i$, is insensitive to the sign choice in~(\ref{Translation-invariant-triads}). However this sign is not entirely unphysical: it represents the \emph{orientation} of our spatial slices. One of the main consequences of the result~\cite{ThroughTheBigBang} is that an orientation flip happens at the singularity, and this has potentially observable physical consequences.

\begin{figure}[t]
\begin{center}
 \includegraphics[width=0.75\textwidth]{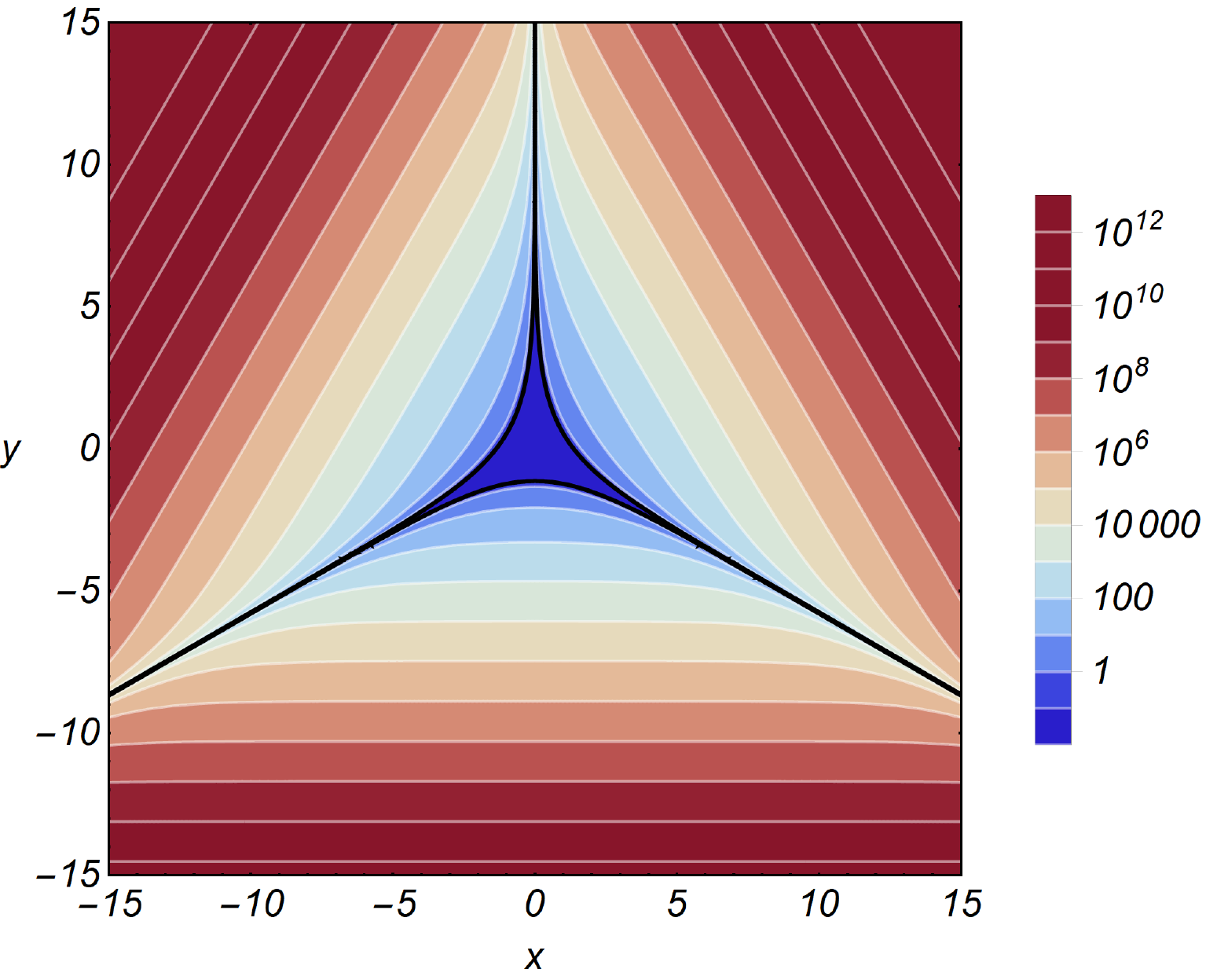}
\end{center}
\caption{Shape potential of the Bianchi IX model. The black curve is where the potential is zero. Inside it is negative  (and it reaches its absolute minimum at $U(0,0) = -3/2$), and outside it is positive.}\label{ShapePotentialDiagram}
\end{figure}

\begin{figure}[t]
\begin{center}
 \includegraphics[width=0.5\textwidth]{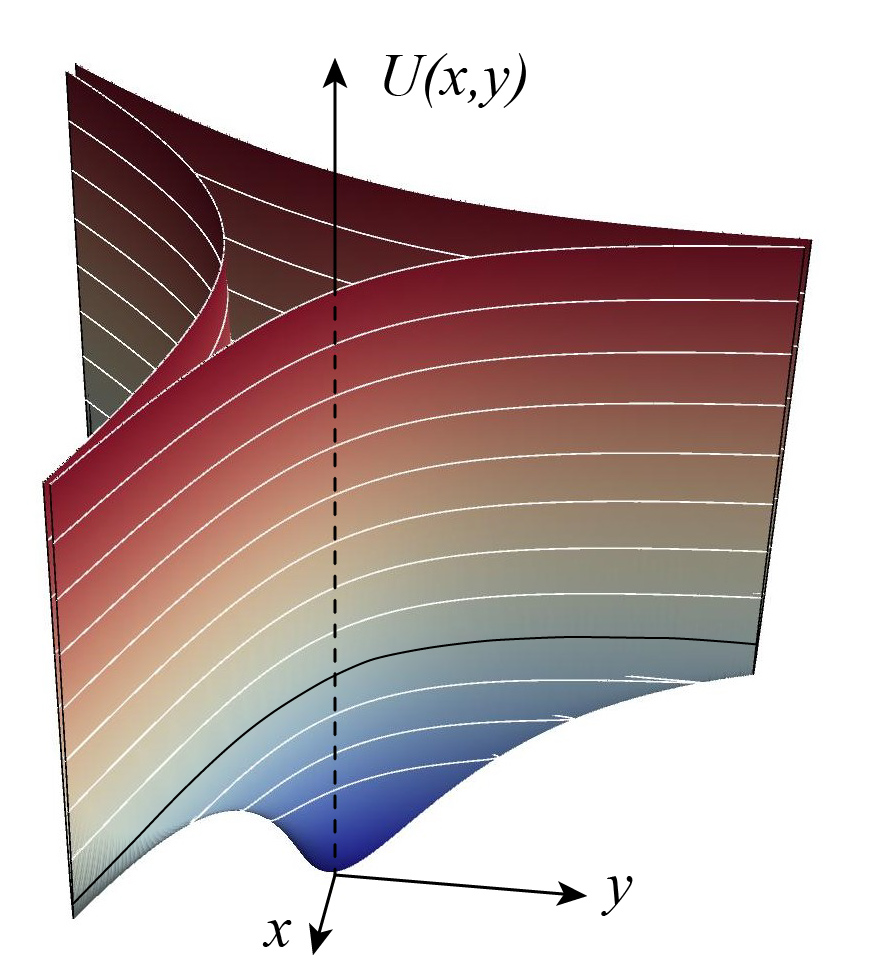}
\end{center}
\caption{Shape potential of the Bianchi IX model in a 3D plot, which highlights the steepness of the potential walls.}\label{ShapePotential3DDiagram}
\end{figure}

As shown in  Appendix~\ref{Appendix_A}, the dynamics of the $x$, $y$, $v$ and $\phi$ variables is described by the following Hamiltonian constraint 
\begin{equation} \label{BIXHamiltonianConstraint_potential}
\mathcal H = {\sfrac 3 8} v^2 \, \tau^2 - \left( k_x^2 + k_y^2 + {\sfrac 1 2} \pi_\phi^2
 \right) - v^{4/3} \, U(x,y) - v^2 \, V(\phi)  \approx 0 \,,
\end{equation}
where $U(x,y)$ is the \emph{shape potential}, defined as:
\begin{equation}\label{ShapePotential}
 U(x,y) :=  f(-2y)+f(\sqrt 3 x +y)+f(-\sqrt 3 x + y)  \,, \qquad
f(x) := e^{- z/\sqrt{6}} - {\sfrac 1 2} e^{2z/\sqrt{6}}\,.
\end{equation}
I have plotted the shape potential in Figs.~\ref{ShapePotentialDiagram} and~\ref{ShapePotential3DDiagram}.  The canonical structure is specified by the following symplectic potential
\begin{equation}
\Theta = \tau \, \delta v + k_x \, \delta x + k_y \, \delta y + \pi_\phi \, \delta \phi \,.
\end{equation}

The kinetic part of the Hamiltonian constraint can be transformed into the Hamiltonian for a free relativistic point particle in 4D Minkowski spacetime. If we make the canonical transformation
\begin{equation}
v = v_0 \, e^{- \frac{\sqrt{3}}{2}  x^0} \,, \qquad p_0 =  \sqrt{{\sfrac 3 4}} v \, \tau \,,
\end{equation}
the two new variables $x^0$ and $p_0$ are canonically conjugate, $\{ x^0,p_0 \} =1$.
Moreover, if we make the trivial change of coordinates
\begin{equation}
x^1 = \phi \,, \qquad x^2 =  {\sfrac 1 {\sqrt{2}}} \, x \,, \qquad x^3 =  {\sfrac 1 {\sqrt{2}}} \, y \,, 
\end{equation}
the canonically conjugate momenta are then
\begin{equation}
p_1 = \pi_\phi \,, \qquad p_2 =\sqrt{2} \, k_x \,, \qquad p_3 = \sqrt{2}  \, k_y \,, 
\end{equation}
and the Hamiltonian constraint becomes
\begin{equation} \label{BIXHamiltonianConstraint3}
\begin{aligned}
&\mathcal H = {\sfrac 1 2} \left( p_0^2 - p_1^2 - p_2^2 - p_3 ^2 \right) - W(v,x) \approx 0 \,,\\
&W(v,x) =  v_0^{\frac 4 3}  \, e^{- \frac{2}{\sqrt{3}} x^0} \, U(x^2,x^3) + v_0^2 e^{-  \sqrt{3} x^0} \, V(x^1)  \,.
\end{aligned}
\end{equation}

\subsection{Kasner epochs}

The shape potential $U(x,y)$ is extremely steep, as is illustrated by the 3D plot of Fig.~\ref{ShapePotential3DDiagram}. The consequence of this is that the motion in such a potential is similar to that of a billiard ball in a triangular pool table: long periods of uniform straight motion (called Kasner epochs), interrupted by sudden violent scatters off the potential walls (called Taub transitions). The chaotic nature of such motion led Misner to call this  a `mixmaster' system~\cite{MTW}.

When the representative point is sufficiently far from the potential walls of $W(v,x)$, the motion is well described by a Kasner (Bianchi I-type)  solution, which is generated by the  Hamiltonian
\begin{equation} \label{BIHamiltonianConstraint1}
\mathcal H_\text{Kasner} = {\sfrac 1 2} \left(p_0^2 - p_1^2 - p_2^2 - p_3 ^2 \right)   \approx 0 \,
\end{equation}
This is just the Hamiltonian constraint of a massless point particle in Minkowski spacetime. The equations of motion are
\begin{equation}
\dot x^\mu = \eta^{\mu\nu} p_\nu \,, \qquad \dot p_\mu = 0\,,
\end{equation}
and their solutions are
\begin{equation}\label{BIsolution1}
x^\mu = \eta^{\mu\nu} p_\nu^0 \, t + x^\mu_0 \,,
 \qquad  p_\mu = p_\mu^0 \,,
\end{equation}
where $\eta^{\mu\nu} = \text{diag} (-1,+1,+1,+1)$, plus the constraint $p_0^2 = p_1^2+p_2^2+p_3^2$.
The coordinates $x^\mu$ evolve linearly in $t$, and therefore the triads~(\ref{Translation-invariant-triads}) will grow or decrease exponentially:
\begin{equation}
e^1 \propto\, e^{\left(- \frac{p_0}{2 \sqrt 3 } - \frac{p_2}{2\sqrt 2} + \frac{p_3}{2\sqrt 6 }\right) t }  \sigma^1 \,,  ~~~
e^2 \propto e^{\left(- \frac{p_0}{2 \sqrt 3 } + \frac{p_2}{2\sqrt 2} + \frac{p_3}{2\sqrt 6 }\right) t  }  \sigma^2  \,, ~~~
e^3 \propto e^{\left(- \frac{p_0}{2 \sqrt 3 } - \frac{p_3}{\sqrt 6} \right) t  }   \sigma^3 \,.
\end{equation}
The exponents in the last equation sum up to $- \frac{ \sqrt 3 p_0}{2}$, so, if we choose a positive value for $p_0$ (the other choice corresponds to the time-reversed solution), the volume will always be contracting throughout the solution. Moreover, if the scalar field momentum $p_1=0$, then the Hamiltonian constraint $p_0 = \sqrt{ p_2^2 + p_3^2}$ implies that one exponent is positive while two are negative, whatever the values of $p_2$ and $p_3$. So there are always two contracting and one expanding direction. If $p_1 \neq 0$ there will be directions in the $p_2,p_3$ plane in which all three directions are contracting, and if $|p_1|$ is sufficiently large there will be only contracting directions.


\begin{figure}[t!]\center
\includegraphics[width=0.5\textwidth]{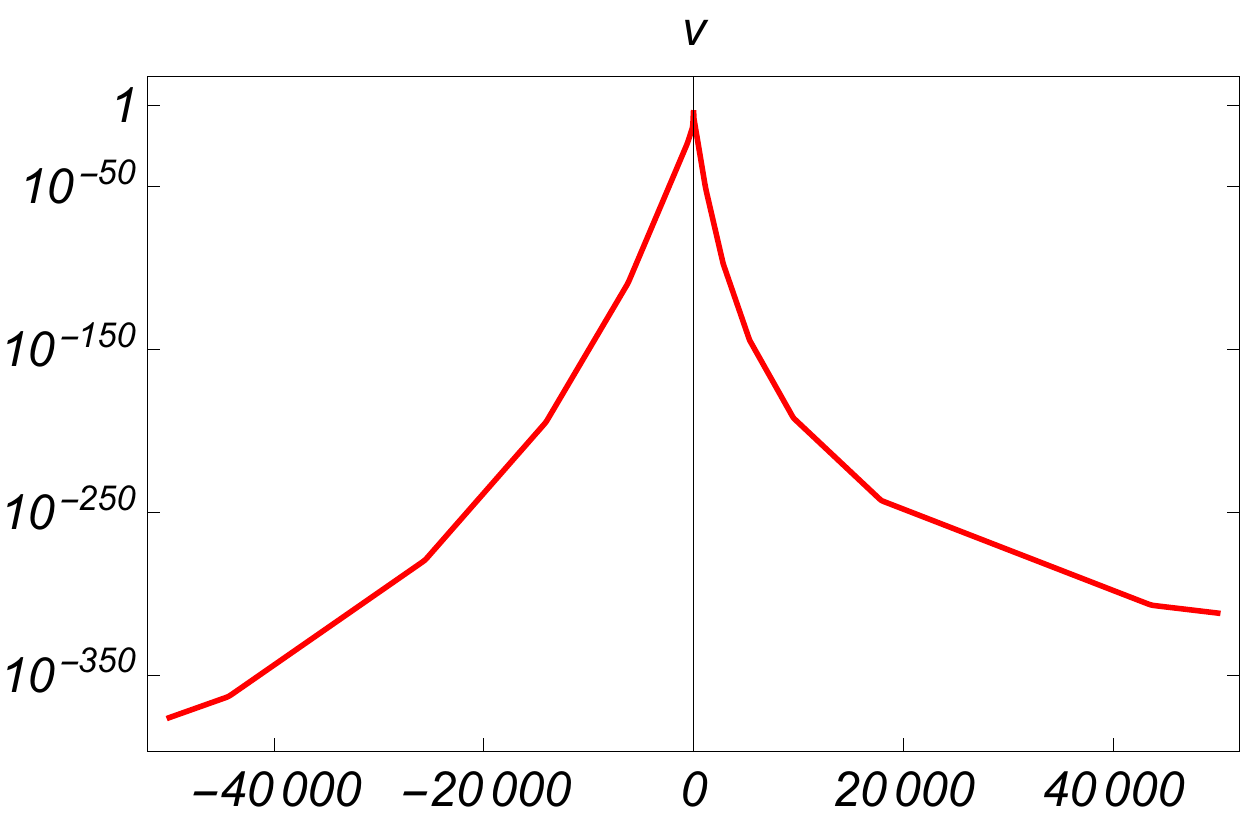}
\caption{Logarithmic plot of the volume in one Bianchi IX solution (calculated numerically).\label{VolumePlot}}
\end{figure}

\subsection{Mixmaster behaviour}

We described the Bianchi IX dynamics as that of a billiard ball in a pool table, but this is slightly inaccurate because the potential walls are actually time-dependent. In fact, the potential $U(x,y)$ is multiplied, in the Hamiltonian constraint~(\ref{BIXHamiltonianConstraint_potential}), by a factor $v^{4/3}$, which is dynamically changing. 

As we have seen, during a Kasner epoch the volume  evolves monotonically. This is true in general for a whole half of each Bianchi IX solution. In fact it is easy to prove that the inverse of the volume has a positive-definite time derivative: $\frac{\d^2 (v^{-1})}{\d t^2} = \frac 3 {16} v \tau^2 + k_x^2 + k_y^2 >0$~\cite{FlavioSDbook} (we are still assuming $p_1=V(\phi)=0$ for the moment). Then $v^{-1}$ is a U-shaped function, with a single minimum and growing monotonically to infinity in the two time directions away from it. Consequently, $v$ has a maximum (at the point of recollapse, when the York time $\tau=0$), and grows monotonically to zero away from it\footnote{This two-branched behaviour can be stopped, for example, if we introduce a sufficiently large positive cosmological constant, which will prevent recollapse and make the volume grow unboundedly in one time direction. Then one has a monotonic volume all the way, going from a big bang singularity at one time asymptote and to infinity at the opposite end.} (see, for example, the simulation of Fig.~\ref{VolumePlot}). The important point is that the exponential dependence of $v$ on $t$ during Kasner epochs implies that it can only go to zero in an infinite amount of parameter time $t$. However, it is not hard to prove that the proper time between any finite $t$ and $t \to \pm \infty$ integrates to a finite value~\cite{FlavioSDbook}, so the big bang singularity is at a finite proper time, and the singularity is a genuine one.

\begin{figure}[t!]\center
\scalebox{1.07}{\includegraphics[width=0.308\textwidth]{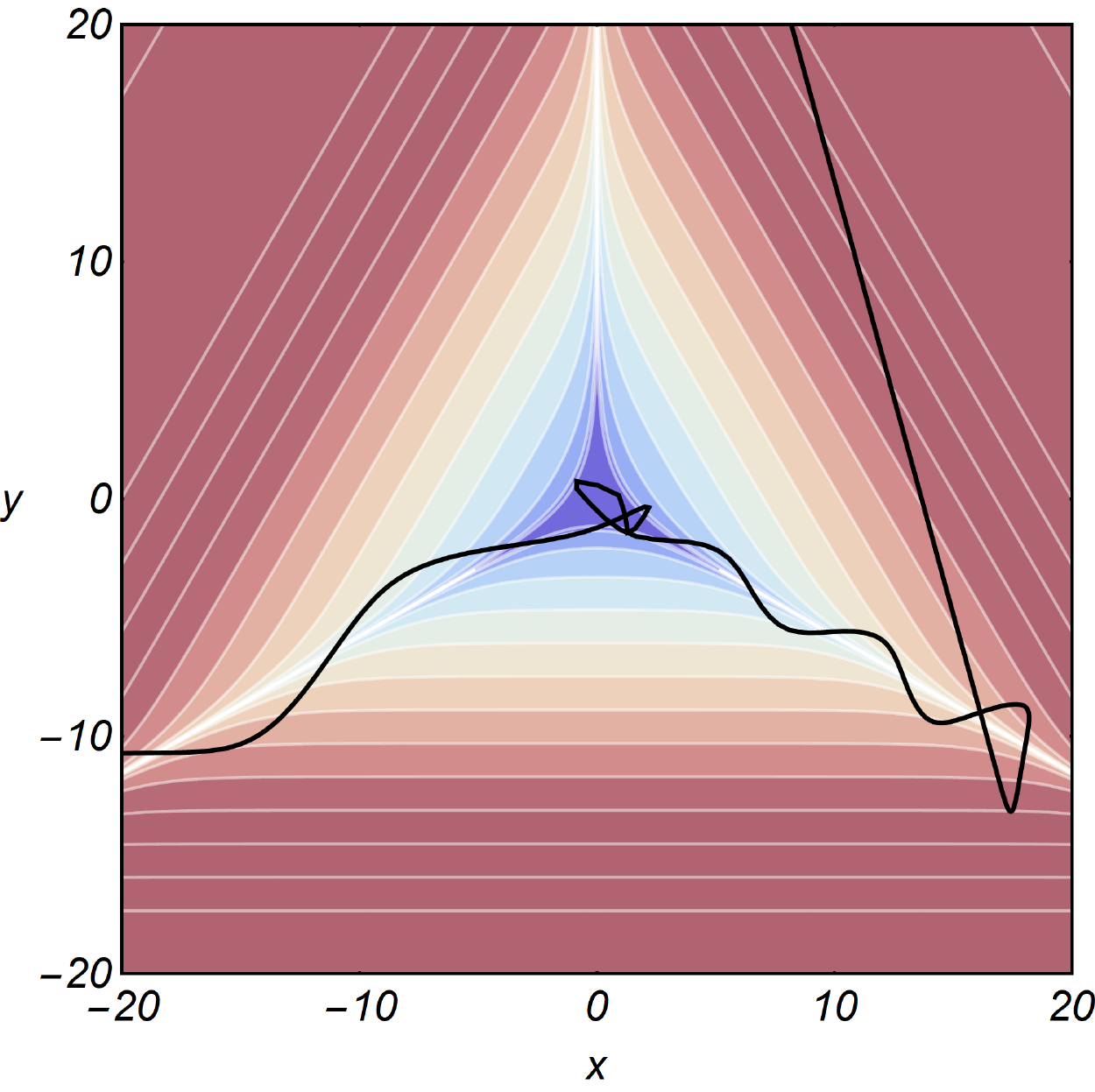}
\includegraphics[width=0.3\textwidth]{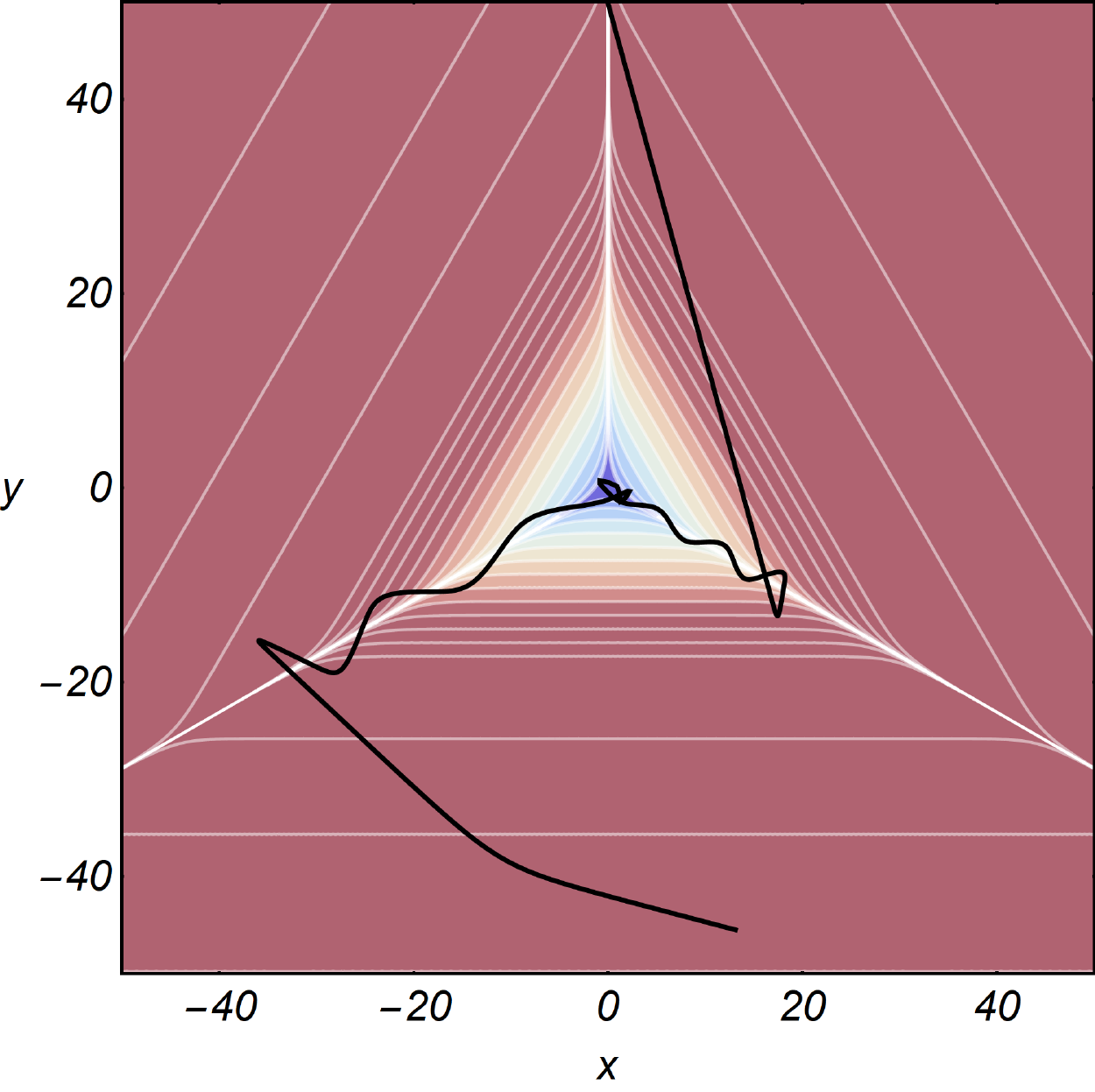}
\includegraphics[width=0.315\textwidth]{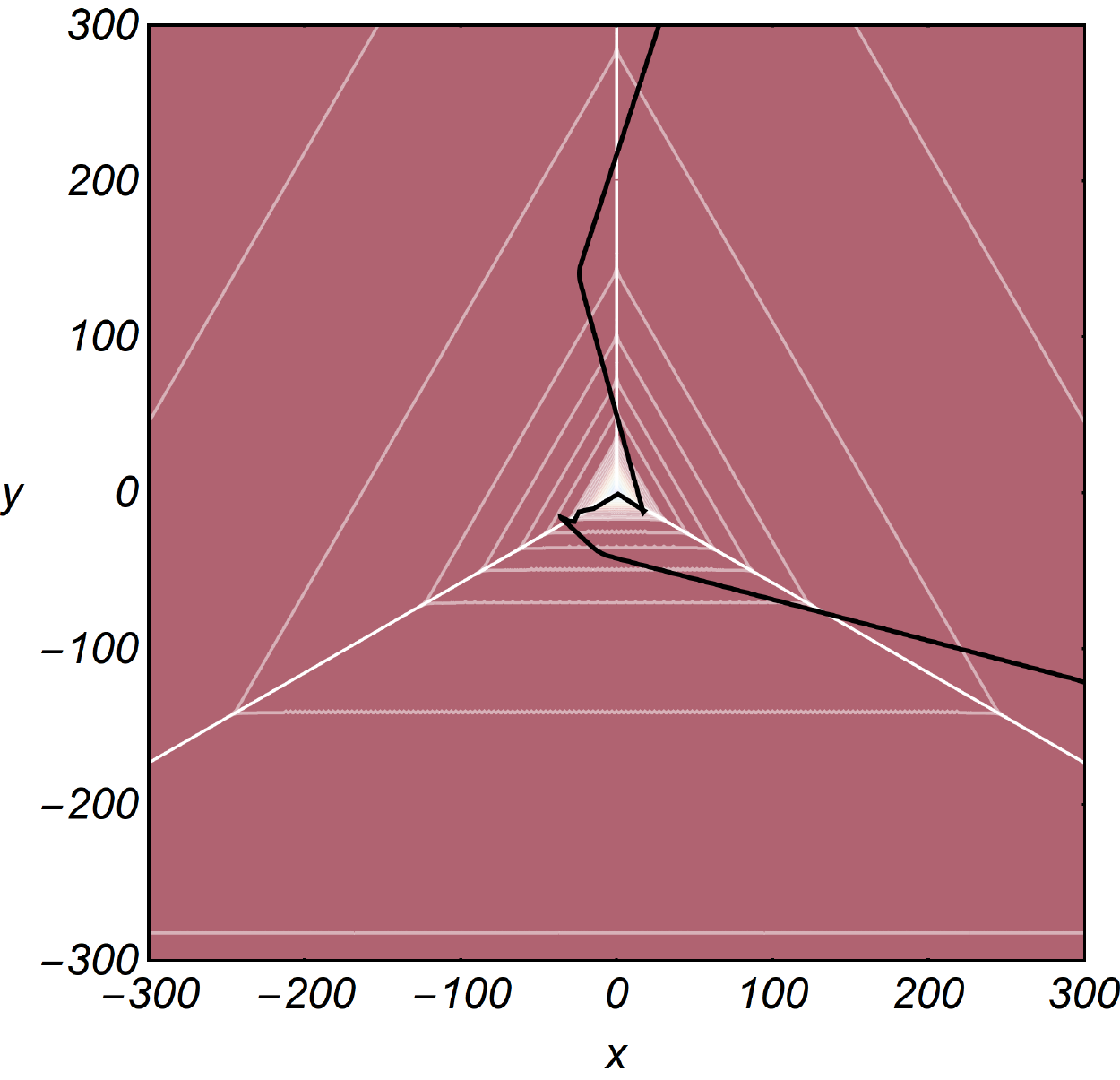}}
\caption{Numerical simulation of the same Bianchi IX solution, integrated over an increasingly large time interval. The solution is seen to explore larger and larger regions of the $(x,y)$ plane.\label{SimulationFig}}
\end{figure}

Coming back to the Bianchi IX potential, its coupling to the volume makes it grow monotonically smaller away from the maximal expansion point. This means that along each Kasner epoch, during which the shape kinetic energy $K =k_x^2 + k_y^2 = \frac 1 2 \left(p_2^2 + p_3^2 \right)$ is conserved, the equipotential line of $U(x,y)$ that corresponds to that value of $K$ will keep changing, in particular shifting towards equipotential lines farther away from the origin.  So, each time the particle scatters of the potential, it will do so at a higher equipotential line. This is like a billiard ball in a triangular pool table \emph{whose walls are moving apart}.  As time passes, the particle explores larger and larger regions of shape space. This can be seen explicitly in Fig.~\ref{SimulationFig}, where I have plotted the same numerical solution over a different time intervals.

\begin{figure}[t!]\center
\includegraphics[width=0.8\textwidth]{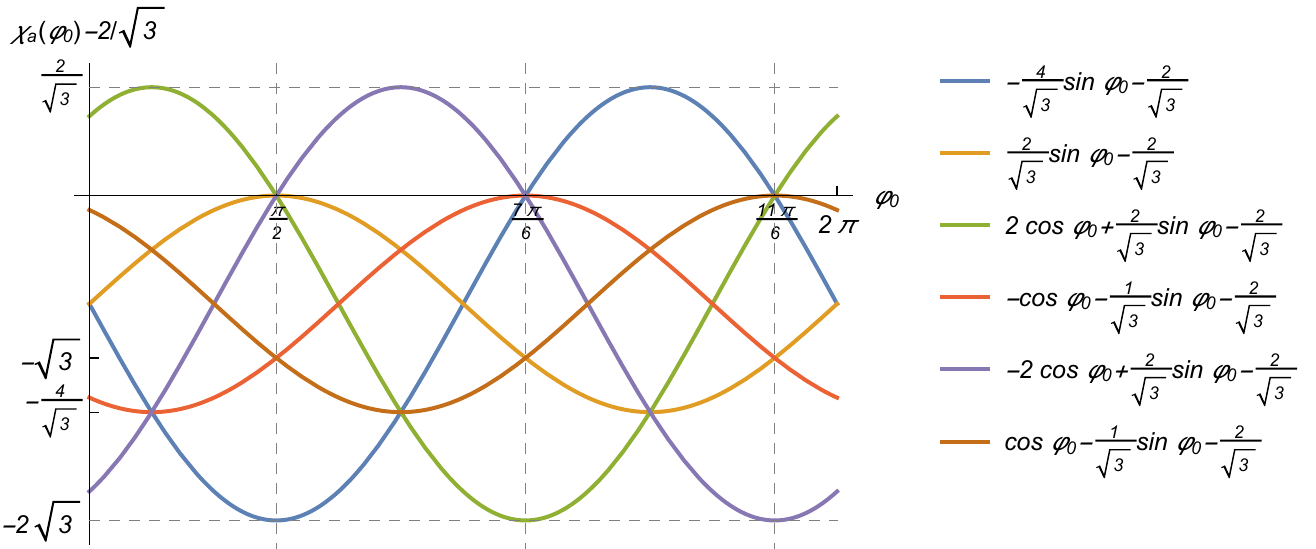}
\label{FigureChi}
\caption{$\chi_a(\varphi_0)- \frac 2 {\sqrt 3}$}
\end{figure}

The issue is now: do the potential walls run away sufficiently fast that the particle never catches up with them? In plain Bianchi IX (no scalar field, or if the scalar field has (exactly) zero momentum and zero potential) this will never happen. Each Kasner epoch  will inevitably end by hitting a wall of the potential $U(x,y)$ and bouncing off of it~\cite{FlavioSDbook}. In fact, the  Bianchi IX potential can be written, in terms of the $x^\mu$ coordinates, as
\begin{equation}
U(x^2,x^3) =  \sum_{a=1}^6 (-1)^a \exp \left(  \sqrt{ (x^2)^2 + (x^3)^2 } \chi_a (\varphi) \right) \,,
\end{equation}
where $\varphi = \arctan x^3/x^2$, and
\begin{equation}
\chi_a =  \textstyle
\{-\frac{4 \sin\varphi }{\sqrt{3}},\frac{2\sin\varphi }{\sqrt{3}},2 \cos\varphi +\frac{2\sin\varphi }{\sqrt{3}},-\cos \varphi -\frac{\sin\varphi }{\sqrt{3}},-2 \cos \varphi +\frac{2\sin\varphi }{\sqrt{3}},\cos \varphi -\frac{\sin\varphi }{\sqrt{3}}\}
 \,.
\end{equation}
Then, if we call $\varphi_0 = \arctan p_3^0/p_2^0$ the direction of the momentum of our Kasner solution~(\ref{BIsolution1}), we get that, for large $t$,
\begin{equation} \label{v43C_BianchiI}
v^{\frac 4 3} C(x^2,x^3)  \to  \text{\it{const.}} \sum_{a=1}^6  \exp\left[ \left( \sqrt{p_2^2+p_3^2} \, \chi_a(\varphi_0)- \frac{2}{\sqrt{3}} p_0 \right) t \right]
\end{equation}
and, from the Hamiltonian constraint in the Kasner regime~(\ref{BIHamiltonianConstraint1}), we know that
\begin{equation}
p_0 = \pm \sqrt{p_1^2 + p_2^2+p_3^2} \,,
\end{equation}
so if $p_1=0$ and for $p_0 > 0$ (shrinking solution) the exponentials in~(\ref{v43C_BianchiI}) are of the form
\begin{equation}\label{Bouncing_condition_1}
\exp\left[ \sqrt{p_2^2+p_3^2} \left( \, \chi_a(\varphi_0) - \frac{2}{\sqrt{3}} \right) t \right] \,.
\end{equation}
Plotting the six functions $\chi_a(\varphi_0) - \frac{2}{\sqrt{3}}$ (Fig.~\ref{FigureChi}), we see that for any value of $\varphi_0$ there is always one function that is positive. This means that one of the six exponentials that form $v^{\frac 4 3} U(x^2,x^3) $ will always be increasing during a Kasner epoch. The potential term then will necessarily catch up with the kinetic one, and the representative point will bounce off the potential walls sooner or later.

The resulting behaviour is that of a billiard ball in a expanding triangular-shaped pool table which never stops bouncing (in parameter time $t$). 
As I said, although the singularity is reached in a finite proper time, the volume goes to zero only as $|t| \to \infty$. It is easy to convince oneself that, in this interval, the shape of the spatial slice goes through an infinite number of Kasner epochs separated by `Taub' bounces. This is what Misner called `mixmaster' behaviour~\cite{MTW}, suggesting that the chaotic nature of this dynamics would make it a perfect scrambler that, as the singularity is approached, erases any information about the initial condition away from it. It is clear that the shape degrees of freedom do not admit a definite limit at the singularity, just like, for example, the function $\sin (1/x)$ as $x \to 0$.

\subsection{Quiescence induced by stiff matter}

If we are in a Kasner epoch (\emph{i.e.} $W(v,x)$ is negligible), but this time the scalar field has a nonvanishing momentum, $p_1 \neq 0$, Eq.~(\ref{v43C_BianchiI}) is changed into
\begin{equation} \label{Bouncing_condition_2}
\exp\left[ \sqrt{p_2^2+p_3^2} \left( \chi_a(\varphi_0) - w\, \frac{2}{\sqrt{3}} \right) t \right] \,, \qquad w =\textstyle \sqrt{1+\frac{p_1^2}{p_2^2+p_3^2}} \,.
\end{equation}
From a look at Fig.~\ref{FigureChi} it is apparent that the functions $\chi_a(\varphi_0) - \frac{2}{\sqrt{3}}$ are negative for certain intervals of $\varphi_0$, and they become negative everywhere for $w \geq 2$. This implies that, if the kinetic energy of the scalar field becomes sufficiently large with respect to $p_2^2+p_3^2$, then no matter in which direction our particle is going, it will stop being influenced by the potential and its trajectory will stabilize around a Kasner solution, without further bounces.

This is what is referred to as `quiescence', and it is an inevitable consequence of coupling Bianchi IX with a free scalar field (\emph{i.e.} $V(x^1)=0$). In fact, a free scalar field is uninfluenced by the dynamics of the geometry, and its evolution is simply that of a point particle moving inertially with conserved momentum: $\phi = x^1 =p_1 \, t + x_1^0$, $\dot p_1 =0$. On the other hand, one can show how the momenta associated with the shape degrees of freedom, $(k_x,k_y)$ [or $(p_2,p_3)$], are (approximatively) conserved only during Kasner epochs, while their bounces against the potential walls of $v^{4/3}U(x^2,x^3)$ are inelastic. This is because the volume that multiplies $U(x^2,x^3)$  in the equations of motion is changing, and if we are evolving towards the singularity it is decreasing. In that case the `shape kinetic energy' $\frac 1 2( p_2^2 + p_3^2)$ cannot be conserved throughout a bounce: since, in the time the particle takes to climb up and back down the potential wall, the potential has decreased, the `downhill' part of the motion will be on a slightly lower potential than the one the particle climbed up on. So the  shape kinetic energy slightly decreases after each bounce (in the time direction of collapse), and eventually it will end up
being dominated by the conserved scalar field kinetic energy $\frac 1 2 p_1^2$. So the quantity $w =\textstyle \sqrt{1+\frac{p_1^2}{p_2^2+p_3^2}} $ in Eq.~(\ref{Bouncing_condition_2}) will eventually grow sufficiently large that quiescence will be attained.

\subsection{Quiescence in presence of a potential for the scalar field}

 If the scalar field has a potential, things can change significantly. For example, if the scalar potential grows sufficiently fast, the existence of quiescent solutions can be excluded. Consider, for instance, a scalar field potential of the form
\begin{equation}\label{Nonquiescent_potential}
V(x^1) = e^{c \,  (x^1)^{1+\epsilon}} \,,
\end{equation}
with $c$ and $\epsilon$ some real positive numbers. If we set up our initial conditions so that we begin in a Kasner epoch, \emph{i.e.}
\begin{equation}
\frac 1 2 \left( p_1^2 +p_2^2 +p_3^2 \right) \gg W(v,x) \,,
\end{equation}
and moreover we require that the scalar field is ultrarelativistic:
\begin{equation}
\frac 1 2  p_1^2   \gg v^2 V(x^1) \,,
\end{equation}
so that its momentum $p_1$ is approximately conserved and we can assume $x^\mu \sim \eta^{\mu\nu} p^0_\nu \, t + x^\mu_0$, $p_0 \sim \sqrt{p_1^2+p_2^2+p_3^2}$, then the potential goes like
\begin{equation}\label{Nonquiescent_potential_limit}
v^2 V(x^1) \propto  e^{-\frac{\sqrt 3}{2}\sqrt{p_1^2+p_2^2+p_3^2} t + c \, (p^1 \, t )^{1+\epsilon} } \,,
\end{equation}
which, for sufficiently large values of $t$, grows monotonically. This excludes the existence of quiescent solutions in the case of potentials that grow as fast as Eq.~(\ref{Nonquiescent_potential}) or faster.

\begin{figure}[t]
\center
\includegraphics[width=0.7\textwidth]{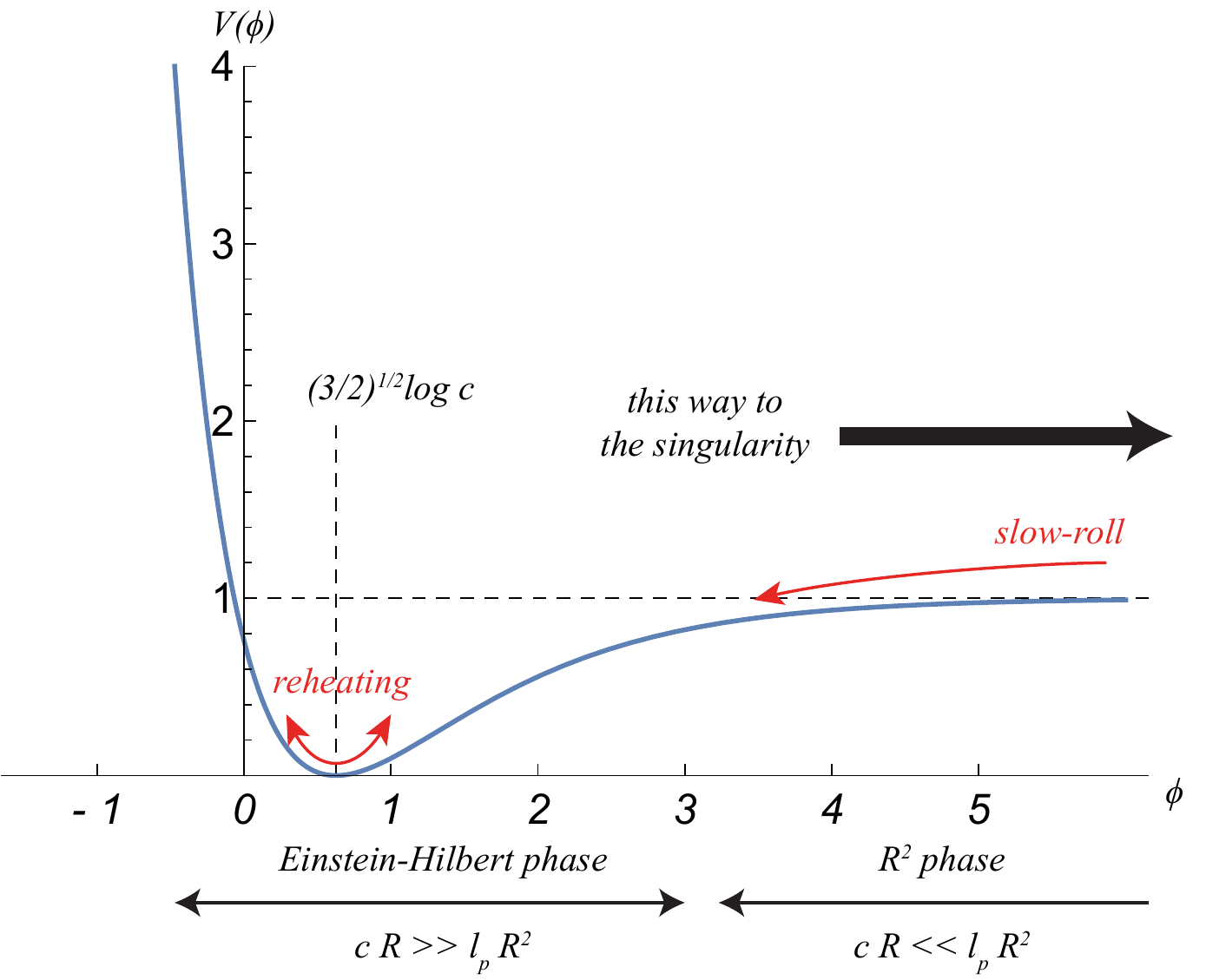}
\label{StarobinskiPotential}
\caption{The scalar field potential of Starobinsky's inflationary model. If we `play backwards' an inflationary solution towards the Big Bang, the scalar field will be seen to climb up the slow-roll plateau with an increasingly well-conserved momentum. The evolution of the scalar field will stabilize around a free-field solution in which the field value $x^1$ grows linearly in parameter time $t$. If the conditions for quiescence are met (among which is $p_1 \neq 0$), the evolution of the shape degrees of freedom $x^2$, $x^3$ and the scalar field $x^1$ will proceed homogeneously on a straight line in the $(x^1,x^2,x^3)$ space all the way to the boundary where $x^1$ and $(x^1)^2+(x^2)^2$ are infinite.}
\end{figure}

In the general case, however, the existence of quiescent solutions cannot be excluded: it is sufficient to start from a sufficiently `ultrarelativistic' initial condition [$p_1^2 \gg v^2 V(x^1)$] in a shrinking Kasner epoch, and the damping factor $v^2$ can overpower most commonly-considered potentials, in particular a mass term $V(x^1) = {\sfrac 1 2} (x^1)^2$ or inflationary potentials in a slow-roll phase. I am particularly interested in the latter case, as the main goal of the present paper is to extend the previous result~\cite{ThroughTheBigBang} of deterministic continuation through the Big Bang to more realistic cosmological models. The most emblematic choice of potential is perhaps Starobinsky's~\cite{Starobinsky1979,Starobinsky1980}
\begin{equation}
V(x^1) = \left(1- c \, e^{-\sqrt{\frac{2}{3}} x^1 }\right)^2 \,,
\end{equation} 
which has several attractive features. First of all, current cosmological constraints place Starobinsky's potential right in the middle of the allowed window~\cite{Planck2018-X}. Secondly, Einstein gravity coupled to a scalar field with that potential is equivalent to a theory with an $R^2$ correction added to the Einstein--Hilbert action. Such a theory implements the first-order quantum corrections to Einstein gravity~\cite{Calmet2019}, and seems therefore to be strongly supported from what we know of quantum field theory and effective field theories.

Starobinsky's potential tends to a constant as $x^1 \to \infty$, and this is clearly enough to allow for quiescent solutions. In particular, any shrinking solution that starts with sufficiently large positive $p^1$ will be quiescent. It has already been noticed in the literature how Starobinsky's model leads to quiescent solutions~\cite{Capozziello2014}.

\section{Continuation through the singularity}

\subsection{Compactification of shape space}

\begin{figure}[b!]
\begin{center}
\includegraphics[width=0.7\textwidth]{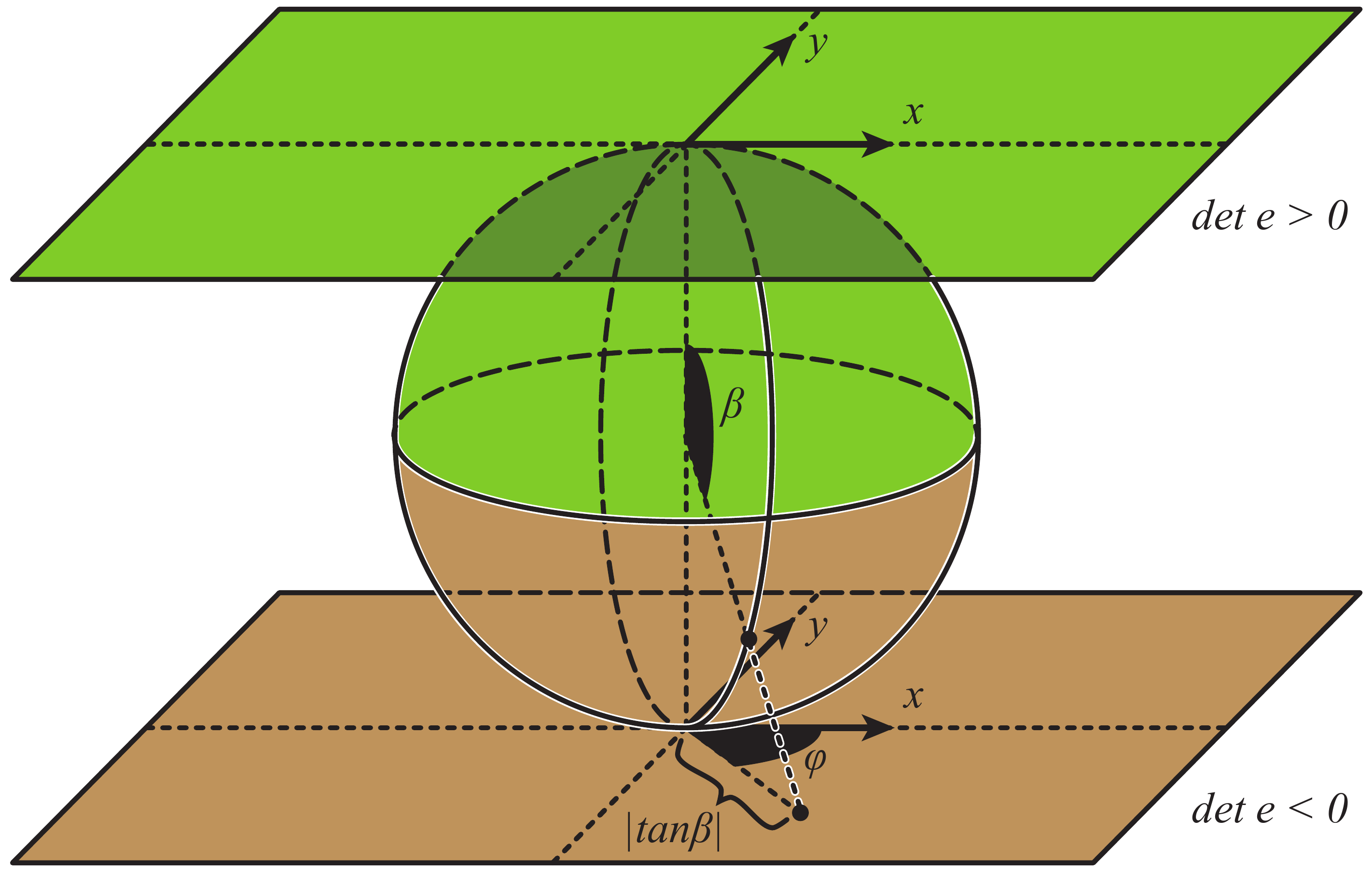}
\caption{Gnomonic projection of a sphere onto the two tangent planes to the poles.}\label{GnomonicFig}
\end{center}
\end{figure}

The variables we used so far to describe the conformal geometry of spatial slices, $(x^2,x^3) \in \mathbbm{R}^2$, can only describe a fixed orientation. This is apparent in Eq.~(\ref{Translation-invariant-triads}), which defines the triads in terms of Misner variables. The sign is a convention, and once it is chosen there is no value that the variables $x^2$ and $x^3$ can take that will change that sign. In order to describe both orientations, we have to extend our configuration space to  the disjoint union $\mathbbm{R}^2 \cup \mathbbm{R}^2$. Now, we have some freedom in choosing the topology of this extended configuration space, which can be stated in terms of ways to glue the boundaries of our two fixed-orientation $\mathbbm{R}^2$ sheets. The rest of the present paper will focus on showing the consequences of a particular choice of topology, first introduced in~\cite{ThroughTheBigBang}, which allows continuation of quiescent solutions in a unique way through the boundary between the two orientations. This topology is that induced by the \emph{gnomonic projection} of our two fixed-orientation planes onto a sphere.

The gnomonic projection is obtained in the following way: represent the two $\mathbbm{R}^2$ planes as parallel planes in an $\mathbbm{R}^3$ ambient space, such that the origin of the $(x^2,x^3)$ coordinate systems of the two planes lie right above each other. Then  consider the sphere that is tangent to these two points. We can now map each point of the two planes into the intersection of the line connecting it to the centre of the sphere and the sphere itself. This way we are mapping each plane onto one of the hemispheres of the sphere (see Fig.~\ref{GnomonicFig}).

\begin{figure}[t!]
\begin{center}
\includegraphics[width=\textwidth]{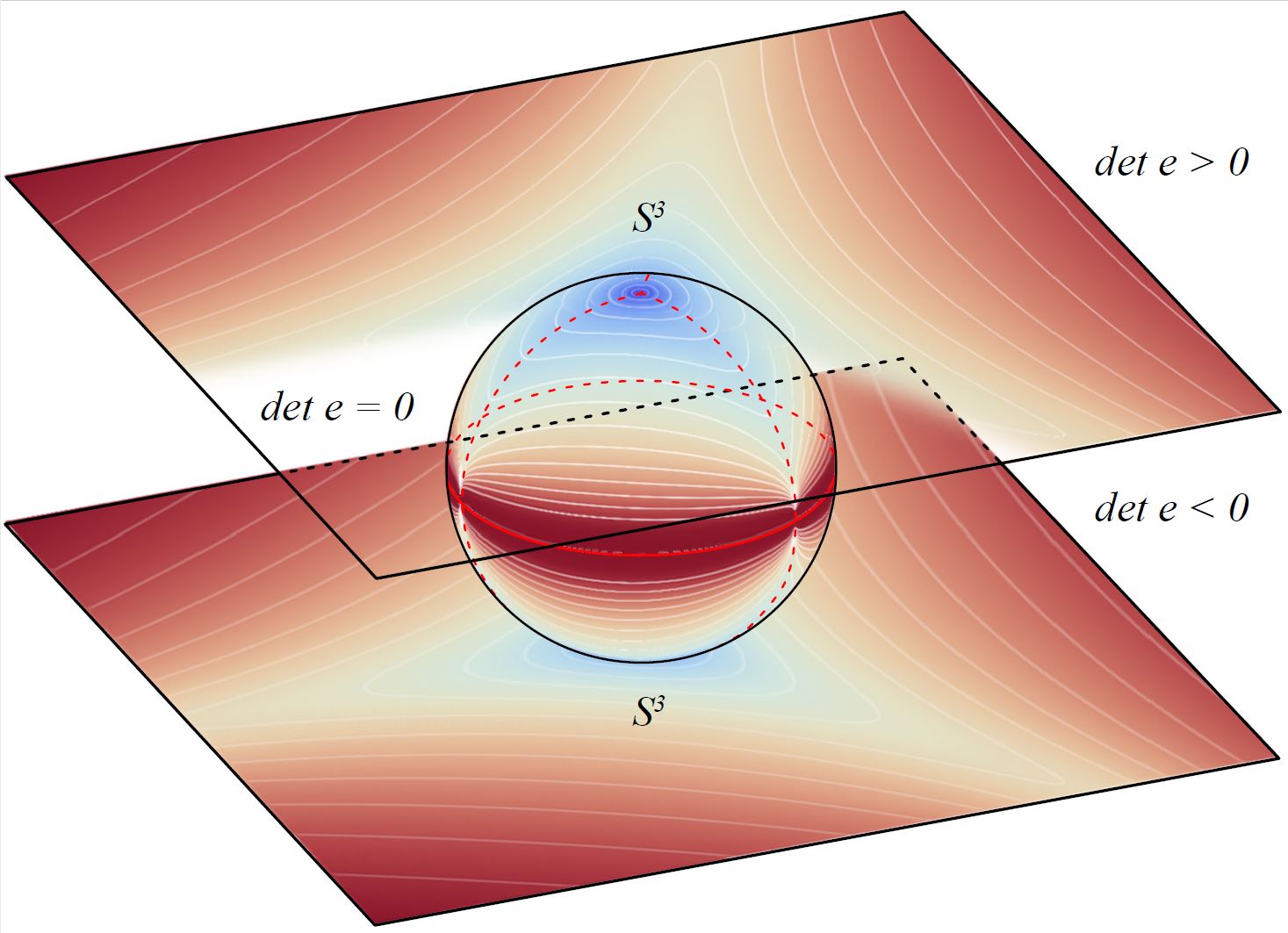}
\caption[Spherical Bianchi IX shape space]{Spherical representation of shape space, with round $S^3$ geometries on the poles and degenerate geometries at the equator. The dashed lines represent geometries with an additional symmetry between two directions and correspond to $\varphi = \pi/6, \pi/2, 5\pi/6$.
The two hemispheres correspond to opposite orientations. The shape potential $U(\varphi,\beta)$ is represented as a colour plot on the sphere with equipotential lines in white.}\label{CompactifiedShapeSpace}
\end{center}
\end{figure}

Coordinatizing the two fixed-orientation planes with polar coordinates, $r = \sqrt{(x^2)^2+(x^3)^2}$ and $\varphi = \arctan (x^3/x^2)$, we have that the radial coordinate corresponds to (the absolute value of) the tangent of the polar angle on the sphere, $r=| \tan \beta|$, while the angle $\varphi$ corresponds to the azimuthal angle on the sphere. Now, at the two poles we have the two isotropic conformal geometries (the 3-sphere) with the two possible orientations, while the equator (where $\beta \to \pi/2$ and $r = |\tan \beta| \to \infty$) corresponds to the border between the two orientations, that is, degenerate geometries in which one of the triads gets infinitely larger than the other two,\footnote{Or, for three isolated choices of $\varphi$, two triads are identical and infinitely larger than the third.} which, in the limit $t\to \infty$, corresponds to an effectively one-dimensional conformal geometry. Such conformal geometries cannot support a nonzero volume (because they are not really three-dimensional), and, as it turns out, the singularity $v\to0$ can only happen as the conformal geometry asymptotes to one of these degenerate configurations. This is the true origin of the Big Bang singularity in this model: a transition to a lower-dimensional conformal geometry. A curve that crosses the equator of our `shape sphere' will correspond to a collapsing geometry that gets squeezed into a one-(or two-)dimensional, zero-volume shape for an instant, and then re-emerges with a finite (and growing) volume and a non-degenerate shape with opposite orientation.
A representation of the potential $U(x^2,x^3)$ on the `shape sphere' is given in Fig.~\ref{CompactifiedShapeSpace}.

\subsection{Kasner epochs on the shape sphere and continuation}

How does a Kasner epoch (a straight line on one of the $(x^2,x^3)$ planes) look on the `shape sphere'?  A point on the sphere in embedding coordinates  $x^2+y^2+z^2=1$ corresponds to the following point on one of the planes: 
\begin{equation} \label{GnomonicProjection}
x^2 = \frac{x}{|z|} \,, \qquad x^3 = \frac{y}{|z|}  \,,
\end{equation}
Writing the sphere in spherical coordinates:
\begin{equation}
\left\{
\begin{aligned}
x &= \sin \beta \, \cos \varphi 
\\
y &= \sin \beta \, \sin \varphi 
\\
z &= \cos \beta 
\end{aligned}
\right. \,,
\end{equation}
we can invert the relations~(\ref{GnomonicProjection}) as:
\begin{equation}\label{SphericalShapeCoordinates} 
x^2  =  |\tan\beta| \, \cos\varphi   \,,
\qquad
x^3  =  |\tan\beta| \, \sin \varphi  \,,
\end{equation}
where, by definition, $\textrm{sign} (\det e) = \textrm{sign} (\tan \beta) = s$.

\begin{figure}[h!]
\begin{center}
\includegraphics[width=\textwidth]{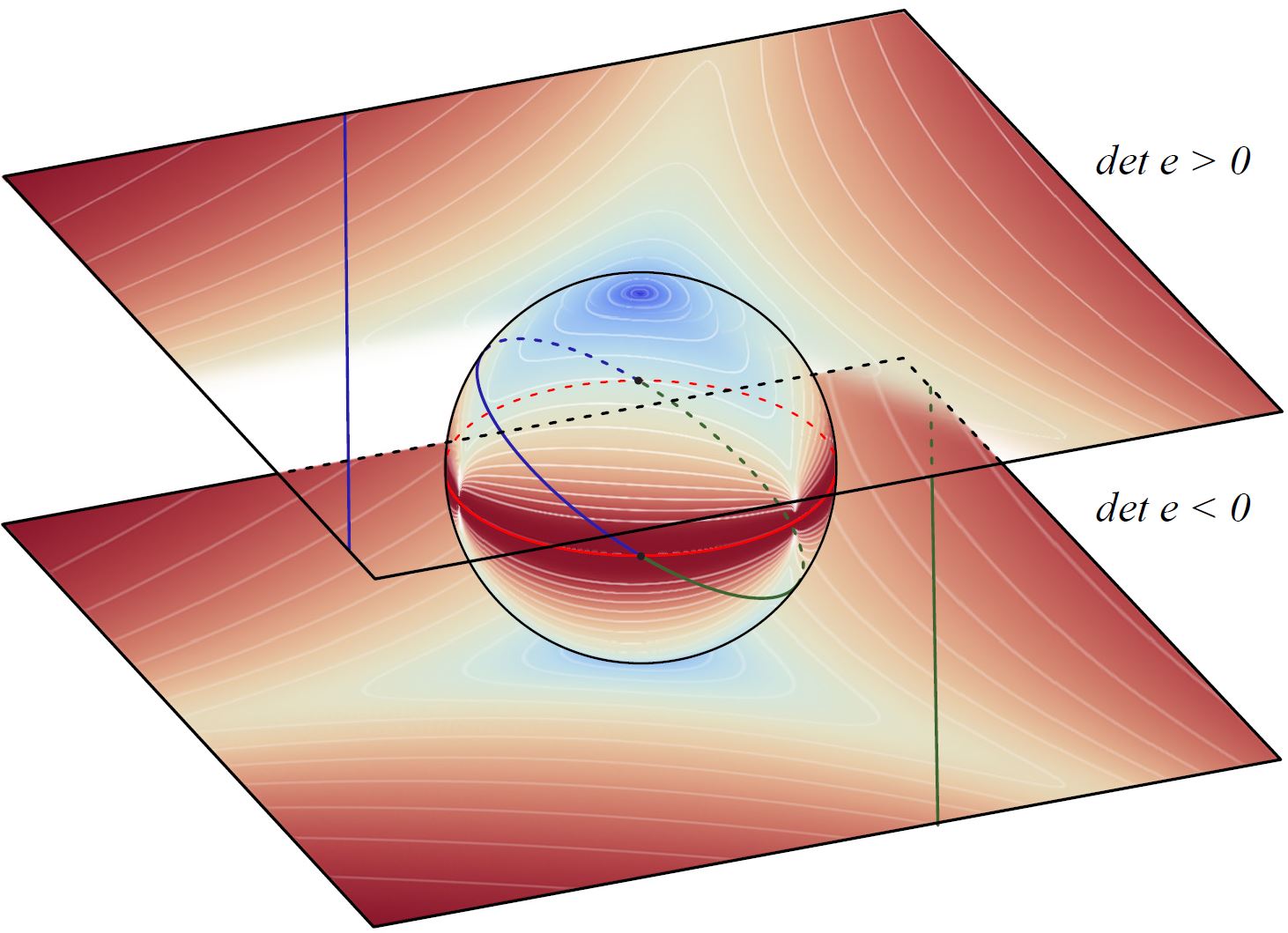}
\caption[Spherical Bianchi IX shape space]{Two Kasner solutions with different orientations, represented as straight lines on the Misner variable planes, and their gnomonic projection onto the shape sphere. These two particular solutions complete a great circle on the shape sphere, and can be considered the continuation of each other through a singularity (at the equator).}\label{CompactifiedShapeSpaceSolution}
\end{center}
\end{figure}

In the coordinates~(\ref{SphericalShapeCoordinates}), a Kasner  solution is mapped into half of a great circle. In fact, these solutions are straight lines in Misner coordinates, which means that they can be represented as solutions of a linear equation:
\begin{equation}
a \, x^2 + b \, x^3 = c \,.
\end{equation}
Using the gnomonic projection rules~(\ref{GnomonicProjection}), we can map this equation into 
\begin{equation}
a \, x + b \, y = c |z| \,,
\end{equation}
which is the equation for two half planes through the origin of 3D space reflected at the $z=0$ plane. Intersecting this with the hemisphere of $x^2 +y^2 +z^2=1$ that corresponds to the chosen orientation, we get half of a great circle.

It is now obvious that, if we want to continue a Kasner solution past the singularity, we simply have to complete the great circle. This corresponds to considering the intersection of the sphere with a plane through the origin:
\begin{equation}\label{EquationOfPlaneIn3D}
a \, x + b \, y + c z  =0 \,, \qquad  a^2 + b^2 >0 \,,
\end{equation}
where the condition $a^2 + b^2 >0$ ensures that the plane is not perpendicular to the $z$-axis (which would make it intersect with the sphere only at the equator).  The projection of this great circle onto the two fixed-orientation planes is simply given by the intersection of~(\ref{EquationOfPlaneIn3D}) with said planes, \emph{i.e.} the equation $a \, x^2 + b \, x^3 \pm c =0$. We see that the two halves of such a solution are related to each other by a parity conjugation around the origin $x^2 = x^3 = 0$. In other words, each solution of the form
\begin{equation}
x^2 = p_2 \, t +   x^2_0 \,,
\qquad
x^3 = p_3 \, t +   x^3_0 \,,
\end{equation}
can be continued past the singularity at $t = \infty$ into the following solution:
\begin{equation}
x^2 = - p_2 \, t -   x^2_0 \,,
\qquad
x^3 = - p_3 \, t -   x^3_0 \,,
\end{equation}
The two straight lines above have the same tangent direction, but they differ from each other because, at their closest point to the origin, they are antipodally related.

\subsection{Including a scalar field in the description}

Here my description will differ from that used in~\cite{ThroughTheBigBang}, in which we limited ourselves to using polar coordinates in the two $(x^2,x^3)$ planes and compactifying the radial coordinate, as we have just seen. Doing this, leaves the scalar field variable $x^1 = \phi$ somewhat out of the picture, a choice that is justified in the massless zero-potential case, inasmuch as the scalar field evolves freely (and uniformly in $t$) in that case, and does not really influence the dynamics of the gravitational degrees of freedom (except indirectly by shifting the Hamiltonian by a constant factor, which, as we have seen, determines the onset of quiescence).  Now, however, since I am considering a scalar field with potential, I will need to determine whether the coupling of this potential with the volume (recall that the scalar field potential appears in the Hamiltonian constraint in the combination $v^2 V(\phi)$) is sufficient to asymptotically suppress the potential and allow the equations of motion to tend to the Kasner equations. In order to do so, we will need to include the scalar field coordinate $x^1$ among those that we compactify. Then, when our compactified radial coordinate $\beta$ reaches the equator $\beta = \pi/2$, we are truly reaching the boundary of the $(x^1,x^2,x^3)$ space.
To achieve this, we can simply consider spherical coordinates in the $(x^1,x^2,x^3)$ space:
\begin{equation}
\left\{
\begin{array}{l}
x^1 = r \,   \cos \theta
\,,\\
x^2 = r \,   \sin \theta \, \cos \varphi
\,,\\
x^3 = r\,  \sin \theta \, \sin \varphi
\,.
\end{array}
\right.
\end{equation}
Now the coordinate $\varphi = \arctan (x^3/x^2)$ has the same meaning as before as a pure shape degree of freedom, while the coordinates $\theta$ and $r$ are a combination of matter and geometric degrees of freedom.
To study the dynamics, we need to find the conjugate momenta to the new variables. The symplectic potential takes the form
\begin{equation}
\Theta =  p_\mu \, \d x^\mu = p_0 \, \d x^0 + J \, \d r  + L_\theta \d \theta + L_\phi \d \phi \,.
\end{equation}
Then the expressions of the momenta conjugate to the new variables, in terms of the old phase-space variables, are
\begin{equation}
\left\{
\begin{array}{l}
J = p_1  \cos \theta + p_2 \sin \theta \, \cos \varphi + p_3 \sin \theta \, \sin \varphi
\,,\\
L_\theta = r \left( p_1  \sin \theta - p_2 \cosh \theta \cos \varphi - p_3 \cosh \theta \sin \varphi \right) 
\,,\\
L_\varphi = r  \sin \theta \left( p_2  \sin \varphi - p_3 \cos \varphi \right) 
\,.
\end{array}
\right.
\end{equation}
The $J,L_\theta,L_\varphi$ momenta are canonically conjugate to $r,\theta$ and $\varphi$. So far we just rewrote the kinematics on one $(x^1,x^2,x^3)$ space in spherical coordinates. We would like to describe two copies of such space, one for each spatial orientation, so we generalize the gnomonic projection procedure to a 3-sphere: we embed two copies of the $(x^1,x^2,x^3)$ space (one for each orientation) as parallel hyperplanes inside an ambient $\mathbbm{R}^4$ space, and project these two hyperplanes onto the two hemispheres of a 3-sphere.
This amounts to compactifying the $r$ coordinate above by introducing $r = |\tan \beta|$ and extending the domain of $\beta$ to the whole $[0,\pi]$ interval. $\beta \in [0,\pi/2]$ describes one orientation and $\beta \in [\pi/2,\pi]$ describes the other.

Now we have to extend the symplectic structure over the whole extended phase space. This choice makes the symplectic potential continuous:
\begin{equation}
\Theta =   p_0 \, \d x^0 +  \frac{J}{\cos^2 \beta} \, \d \beta  + L_\theta \d \theta + L_\phi \d \phi \,,
\end{equation}
and differentiating $\Theta$ we get the following symplectic form:
\begin{equation}
\d \Theta =  \d  p_0 \wedge \d x^0 +  \frac{1}{\cos^2 \beta} \d J \wedge \d \beta  + \d L_\theta  \wedge \d \theta + \d L_\phi  \wedge \d \phi \,,
\end{equation}
whose inverse gives the following Poisson brackets:
\begin{equation}\label{Poisson_Brackets}
\{ p_0 , x^0 \} = 1 \,, \qquad \{ J , \beta \} = \cos^2 \beta  \,, \qquad
 \{ L_\theta , \theta \} = 1 \,, \qquad \{ L_\varphi , \varphi \} = 1 \,, 
\end{equation} 
and all the other ones are zero.
In these coordinates Kasner [\emph{i.e.} ignoring the potential term $W(x)$] Hamiltonian constraint takes the form
\begin{equation}\label{BIHamiltonianConstraint2}
\mathcal H_\text{Kasner} = {\sfrac 1 2} p_0^2 - {\sfrac 1 2}  J^2 - {\sfrac 1 2} \cot^2 \beta  \left( L_\theta^2 + \sin^{-2} \theta \,  L_\varphi^2 \right)  \,.
\end{equation}
Now, using the Poisson brackets~(\ref{Poisson_Brackets}) and the Hamiltonian constraint~(\ref{BIHamiltonianConstraint2}), we get the following equations of motion:
\begin{equation}
\begin{aligned}
& \dot x^0 = p_0 \,,
&&
\dot \beta = - \cos^2\beta \, J \,,
\\
&
\dot \theta = - L_\theta \, \cot^2 \beta  \,,
&&
\dot \varphi = - L_\varphi \, \cot^2 \beta    \,\sin^{-2} \theta \,,
\\
& \dot p_0 = 0 \,, 
&&
\dot J  = - \cot^3 \beta \left( L_\theta^2 + \sin^{-2} \theta \, L_\varphi^2 \right)  \,,
\\
& \dot L_\theta = -   \sin^{-3} \theta \cos \theta   \cot^2 \beta  L_\varphi^2 \,, 
&&
\dot L_\varphi = 0 \,.
\end{aligned}
\end{equation}
With $\beta$ as the independent variable:
\begin{equation}
\begin{aligned}
& \frac{\d x^0}{\d \beta} = - \frac{p_0}{\cos^2 \beta \, J}
\,,
\\
&\frac{\d\theta}{\d \beta} =  \frac{L_\theta}{\sin^2 \beta \, J} \,,
\\
& \frac{\d\varphi}{\d \beta} =  \frac{L_\varphi}{\sin^2 \beta \,\sin^2 \theta \,  J} \,,
\end{aligned}
\qquad
\begin{aligned}
& \frac{\d J}{\d \beta} = \frac{\cos \beta}{\sin^3 \beta \, J}  \left( L_\theta^2 + \sin^{-2} \theta \, L_\varphi^2 \right) \,,
\\
& \frac{\d L_\theta}{\d \beta} =   \frac{ \cos \theta }{\sin^2 \beta \, \sin^3 \theta \, J}  L_\varphi^2 \,, \\
& \frac{\d p_0}{\d \beta} = \frac{\d L_\varphi}{\d \beta} = 0 \,
\end{aligned}
\end{equation}
all of the expressions on the right-hand sides of the above are well-behaved as $\beta \to \pi/2$, except $\frac{\d x^0}{\d \beta} = - \frac{p_0}{\cos^2 \beta \, J}$ which diverges. However notice that the (Kasner) Hamiltonian constraint~(\ref{BIHamiltonianConstraint2}) implies that 
\begin{equation}
 p_0^2 = J^2  + {\sfrac 1 2} \cot^2 \beta  \left( L_\theta^2 + \sin^{-2} \theta \,  L_\varphi^2 \right) \xrightarrow[\beta \to (\frac \pi 2)^\pm]{} J^2 \,,
\end{equation}
so $\left| \frac{\d x^0}{\d \beta} \right| = \xrightarrow[\beta \to (\frac \pi 2)^\pm]{}  \frac 1 {\cos^2 \beta}$, which means that $\left| x^0 \right| \to \pm |\tan \beta |$. Moreover, it is easy to prove that the quantity
\begin{equation}
y^0 = x^0 + \tan \beta \frac{p_0}{J} \,,
\end{equation}
will tend to a finite value as $\beta \to (\frac \pi 2)^\pm$. In terms of $y^0$, $p^0$,  $J$, $\theta$, $L_\theta$, $\varphi$, $L_\varphi$ the equations of motion now take the form
\begin{equation}
\begin{aligned}
& \frac{\d y^0}{\d \beta} = - \frac{p_0}{\sin^2 \beta \, J^3}  \left( L_\theta^2 + \sin^{-2} \theta \, L_\varphi^2 \right)
\,,
\\
&\frac{\d\theta}{\d \beta} =  \frac{L_\theta}{\sin^2 \beta \, J} \,,
\\
& \frac{\d\varphi}{\d \beta} =  \frac{L_\varphi}{\sin^2 \beta \,\sin^2 \theta \,  J} \,,
\end{aligned}
\qquad
\begin{aligned}
& \frac{\d J}{\d \beta} = \frac{\cos \beta}{\sin^3 \beta \, J}  \left( L_\theta^2 + \sin^{-2} \theta \, L_\varphi^2 \right) \,,
\\
& \frac{\d L_\theta}{\d \beta} =   \frac{ \cos \theta }{\sin^2 \beta \, \sin^3 \theta \, J}  L_\varphi^2 \,, \\
& \frac{\d p_0}{\d \beta} = \frac{\d L_\varphi}{\d \beta} = 0 \,.
\end{aligned}
\end{equation}
If we want a complete reduction of the system, we can solve the Hamiltonian constraint w.r.t $p_0$ and replace its expression above, eliminating the equation of motion for $p_0$. The result is six equations expressing the evolution of $y^0$, $J$, $\theta$, $L_\theta$, $\varphi$, $L_\varphi$ with respect to $\beta$. We don't necessarily have to make this last step, however, because the Hamiltonian constraint~(\ref{BIHamiltonianConstraint2}) is regular at $\beta \to \pi/2$, and so are all its first and second derivatives (except at the poles $\theta =0$ and $\beta =0$).

The above  system of equations is well-posed at $\beta = \pi/2$. In fact the right-hand-sides of the equations tend to the same finite limits as  $\beta \to (\pi/2)^\pm$:
\begin{equation}\label{BI_limit_1}
\begin{aligned}
 & \frac{\d y^0}{\d \beta} \xrightarrow[\beta \to (\frac \pi 2)^\pm]{} - \frac{p_0}{ J^3}  \left( L_\theta^2 + \sin^{-2} \theta \, L_\varphi^2 \right) \,,
&&
& \frac{\d\theta}{\d \beta} \xrightarrow[\beta \to (\frac \pi 2)^\pm]{} - \frac{L_\theta}{J} 
 \,,
\\
&  \frac{\d\varphi}{\d \beta} \xrightarrow[\beta \to (\frac \pi 2)^\pm]{}  - \frac{L\varphi}{J \,\sin^2 \theta}  
\,,
&&
&  \frac{\d J}{\d \beta} \xrightarrow[\beta \to (\frac \pi 2)^\pm]{} 0
  \,,
\\
&  \frac{\d L_\theta}{\d \beta}  \xrightarrow[\beta \to (\frac \pi 2)^\pm]{} - \frac{\cos \theta  }{J  \sin^3 \theta  } L_\varphi^2  
  \,,
&&
&  \frac{\d L_\varphi}{\d \beta} \xrightarrow[\beta \to (\frac \pi 2)^\pm]{} 0  
   \,,
  \\
 & \frac{\d p_0}{\d \beta} \xrightarrow[\beta \to (\frac \pi 2)^\pm]{} 0  
  \,,
\end{aligned}
\end{equation}
and the same is true for their first derivatives with respect to $y^0$, $\theta$, $\varphi$, $J$, $L_\theta$, $L_\varphi$, $p_0$ and $\beta$:
\begin{equation}\label{BI_limit_2}
\resizebox{0.9\hsize}{!}{$\begin{aligned}
&\left( \begin{array}{c}
\frac{\partial }{\partial y^0}\\\frac{\partial}{\partial \theta}\\  \frac{\partial }{\partial \varphi}  \\  \frac{\partial}{\partial J} \\ \frac{\partial }{\partial L_\theta} \\ \frac{\partial }{\partial L_\varphi} \\\frac{\partial }{\partial p_0} \\\frac{\partial }{\partial \beta}  
\end{array}
\right)
\otimes
 \left( \begin{array}{c}
 \frac{\d y^0}{\d \beta}\\\frac{\d\theta}{\d \beta}\\  \frac{\d\varphi}{\d \beta}  \\  \frac{\d J}{\d \beta} \\ \frac{\d L_\theta}{\d \beta} \\ \frac{\d L_\varphi}{\d \beta} \\\frac{\d p_0}{\d \beta}  
 \end{array}
 \right)^T
\xrightarrow[\beta \to (\frac \pi 2)^\pm]{}
&
&\frac 1 {\sin^2 \theta}
\left(
\begin{array}{ccccccc}
 0 & 0 & 0 & 0 & 0  & 0 & 0\\
 \frac{2 L_\varphi^2 p_0 \cot  \theta }{J^3} & 0 & -\frac{2 L_\varphi \cot  \theta  }{J} & 0 & -\frac{3 L_\varphi^2 \cot  \theta }{\sin\theta J}  & 0 & 0\\
 0 & 0 & 0 & 0 & 0  & 0 & 0\\
 \frac{3 p_0 \left(L_\theta^2 \sin^2 \theta +L_\varphi^2 \right)}{J^4} & -\frac{\sin^2 \theta L_\theta}{J^2} & -\frac{L_\varphi}{J^2} & 0 & -\frac{L_\varphi^2 }{\sin \theta J^2} & 0 & 0 \\
 -\frac{2\sin^2 \theta  L_\theta p_0}{J^3} & \frac{\sin^2 \theta }{J} & 0 & 0 & 0 & 0 & 0 \\
 -\frac{2 L_\varphi p_0 }{J^3} & 0 & \frac{1}{J} & 0 & \frac{2 L_\varphi}{\sin \theta J} & 0 & 0 \\
 -\frac{\sin^2 \theta  L_\theta^2+L_\varphi^2 }{J^3} & 0 & 0 & 0 & 0 & 0 & 0 \\
 0 & 0 & 0 & -\frac{\sin^2 \theta  L_\theta^2+L_\varphi^2  }{J} & 0 & 0 & 0\\
\end{array}
\right)\,.
\end{aligned}$}
\end{equation}

\subsection{Existence and uniqueness of solutions at the singularity}

We are now ready to prove the main theorem: the full Bianchi IX + scalar field equations satisfy the Picard--Lindel\"of theorem at $\beta = \pi/2$ when expressed in terms of the  variables $y^0$, $J$, $\theta$, $L_\theta$, $\varphi$, $L_\varphi$ with $\beta$ used as independent variable and when the conditions to attain quiescence are met:
\begin{equation}
 \chi_a(\varphi_0) < \frac{2}{\sqrt{3}}   \sqrt{1+\frac{p_1^2}{p_2^2+p_3^2}}
\end{equation}
which in the new coordinates reads
\begin{equation}
 \chi_a(\varphi_0) < \frac{2}{\sqrt{3}}   \sqrt{\frac{\sin ^2 \theta  \left(J^2 \tan^2 \beta + L_\theta^2 \right)+L_\varphi^2}{\sin ^2 \theta  \left( L_\theta \cos \theta - J |\tan \beta | \sin \theta \right)^2 + L_\varphi^2}}\,,
\end{equation}
and also 
\begin{equation}
\lim_{s \to \infty} e^{-\frac{\sqrt 3}{2}\sqrt{p_1^2+p_2^2+p_3^2} s} V(p_1 s) =0 \,,
\end{equation}
that is,
\begin{equation}
\lim_{s \to \infty} e^{-\frac{\sqrt 3}{2}\sqrt{J^2 + \frac{L_\theta^2 +\sin^{-2} \theta L_\varphi^2 }{\tan^2 \beta}} s} V\left[ \left( J \cos \theta + L_\theta \sin \theta |\tan \beta|^{-1} \right) s \right] =0 \,.
\end{equation}

Then the full Bianchi IX + scalar equations of motion, which in our variables read
\begin{equation}
\begin{aligned}
  \frac{\d y^0}{\d \beta}  =& - \frac{p_0}{\sin^2 \beta \, J^3}  \left( L_\theta^2 + \sin^{-2} \theta \, L_\varphi^2 \right)
\,,
\\
 \frac{\d\theta}{\d \beta}  =&  \frac{L_\theta}{\sin^2 \beta \, J} \,,
\\
  \frac{\d\varphi}{\d \beta}  =&  \frac{L_\varphi}{\sin^2 \beta \,\sin^2 \theta \,  J} \,,
\\
  \frac{\d J}{\d \beta}  =& \frac{\cos \beta}{\sin^3 \beta \, J}  \left( L_\theta^2 + \sin^{-2} \theta \, L_\varphi^2 \right) 
\\
& - \cos^2  \beta \frac{\partial}{\partial \beta} \left[ e^{- \sqrt{3} (|\tan \beta|p_0/J -y^0) } V(|\tan \beta| \cos \theta)-  e^{- \frac 2{\sqrt{3}} (|\tan \beta|p_0/J -y^0) }C(\varphi, \beta)  \right]  \,,
\\
  \frac{\d L_\theta}{\d \beta}  =&   \frac{ \cos \theta }{\sin^2 \beta \, \sin^3 \theta \, J}  L_\varphi^2 +  e^{- \sqrt{3} (|\tan \beta|p_0/J -y^0) } \frac{\partial V(|\tan \beta| \cos \theta)}{\partial \theta} \,, \\
 \frac{\d L_\varphi}{\d \beta}  =& -  e^{- \frac 2{\sqrt{3}} (|\tan \beta|p_0/J -y^0) }  \frac{\partial C(\varphi, \beta)}{\partial \varphi} \,,
\\
  \frac{\d p_0}{\d \beta}  =& + \sqrt{3} e^{- \sqrt{3} (|\tan \beta|p_0/J -y^0) } V(|\tan \beta| \cos \theta) - \frac 2{\sqrt{3}}  e^{- \frac 2{\sqrt{3}} (|\tan \beta|p_0/J -y^0) }C(\varphi, \beta)   \,,
\end{aligned}
\end{equation}
admit the same limit as Eqs.~(\ref{BI_limit_1}) and~(\ref{BI_limit_2}) as $\beta \to \pi/2$.
This proves the result.

\subsection{Extension to N scalar fields}

Our result can be extended immediately to any number $n$ of scalar fields $\phi_i$, $i=1,\dots, n$, with arbitrary potential $V(\phi_1, \dots , \phi_n)$ as long as the potential admits quiescent solutions, \emph{i.e.} there are runaway solutions in which $v^2$ goes to zero faster than $V$ grows. We will call 
\begin{equation}
\begin{aligned}
&x^i = \phi_i \,,& & p_i = \pi_{\phi_i} \,,
\\
&x^{n+1} =\frac 1 {\sqrt{2}} q^1 \,,& &  x^{n+2} =\frac 1 {\sqrt{2}} q^2 \,,
\\
&p_{n+1} = {\sqrt{2}} \, \pi_1 \,,& & p_{n+2} =  {\sqrt{2}} \, \pi_2 \,.
\end{aligned}
\end{equation}
Then we can introduce hyperspherical coordinates:
\begin{equation}
\left\{
\begin{array}{l}
x^1 =  |\tan \beta|  \,   \cos \theta^1
\,,\\
x^2 = |\tan \beta| \,   \sin \theta^1 \, \cos \theta^2
\,,\\
\vdots\\
x^{n-1} = |\tan \beta|  \,  \sin \theta^1 \dots \sin \theta^{n-1} \, \cos \theta^{n}
\,,\\
x^{n} = |\tan \beta|  \,  \sin \theta^1 \dots \sin \theta^{n-1} \, \sin \theta^{n}
\,,\\
x^{n+1} =  |\tan \beta|  \,  \sin \theta^1 \dots \sin \theta^{n} \, \cos \theta^{n+1}
\,,\\
x^{n+2} = |\tan \beta|  \,  \sin \theta^1 \dots \sin \theta^{n} \, \sin \theta^{n+1}
\,.
\end{array}
\right.
\end{equation} 
where now $\theta^{n+1}=\varphi$, the angular coordinate on shape space.

The above coordinates satisfy $\sum_{i=1}^{n+2} (x^i)^2= \tan^2 \beta$. If we introduce embedding coordinates $y^i$ in a $(n+3)$-dimensional Euclidean space such that $x^i = y^i$ and $y^{n+3}=\pm 1$, then the coordinates $\beta$, $\theta^a$ are hyperspherical coordinates on the $(n+2)$-dimensional unit hypersphere, and each curve on the plane $y^{n+3}=\pm 1$ will be projected onto a curve on the hemisphere $\beta < \pi/2$ or $\beta> \pi/2$. This is a trivial generalization of gnomonic coordinates to the hypersphere. Just like the single-scalar-field case, straight lines on the $y^{n+3} =\pm1$ hyperplanes correspond to geodesics of the $(n+2)$-sphere because they can be written as the intersection of $n-2$ codimension-1 hyperplanes through the origin with the hypersphere.

The momenta $L_a$ conjugate to the $\theta^a$, $a=1,\dots,n+1$, coordinates ($\{ L_a , \theta^b \} = \delta_{ab}$), together with the `dilatational momentum' $J$ such that $\{ J,\beta\}=\cos^2 \beta$, are given by
\begin{equation}
\left( \begin{array}{c}
J
\\
L_1
\\
\vdots
\\
L_{n+1}
\end{array}\right)
= M \cdot
\left( \begin{array}{c}
p_1
\\
\vdots
\\
p_{n+2}
\end{array}\right)\end{equation}
where $M$ is the Jacobian of the transformation to hyperspherical coordinates:
\begin{equation}
M = \left(
\begin{array}{cccc}
\cos \theta^1 &  \sin \theta^1 \, \cos \theta^2 & \dots &  \sin \theta^1 \dots  \sin \theta^{n+1}
\\
|\tan \beta| \sin \theta^1 &  - |\tan \beta| \cos \theta^1 \, \cos \theta^2 & \dots & - |\tan \beta|  \cos \theta^1  \sin \theta^2 \dots  \sin \theta^{n+1}
\\
0 & |\tan \beta|  \sin \theta^1 \, \sin \theta^2 & \dots & - |\tan \beta|  \sin \theta^1 \cos \theta^2 \dots  \sin \theta^{n+1}
\\
\vdots & \ddots & & \vdots
\\
0&  \dots &  0 & |\tan \beta|  \sin \theta^1 \dots  \sin \theta^{n+1}
\\
0&  \dots &  0 & -|\tan \beta|  \sin \theta^1 \dots  \cos \theta^{n+1}
\end{array}
\right).
\end{equation}
The above change of coordinates induces a nontrivial metric tensor on configuration space:
\begin{equation}\label{metric_hypersphere}
g= M \cdot I \cdot M^T =
\left(
\begin{array}{ccccc}
1 & 0 & 0 & \dots & 0
\\
0 & \tan^2 \beta & 0 & \dots & 0
\\
0 & 0 & \tan^2 \beta \, \sin^2 \theta^1 &  \dots & 0
\\
\vdots  &\vdots&\vdots& \ddots & \vdots
\\
0 & 0 & 0 & \dots & \tan^2 \beta \, \sin^2 \theta^1 \dots \sin^2 \theta^{n+1}
\end{array}
\right) \,.
\end{equation}
This allows us to write the Hamiltonian constraint in terms of the new variables. In fact, call $J=L_0$ and, then
\begin{equation}
\delta^{ij} p_i  p_j = g^{AB} L_A L_B \,, ~~~ A,B= 0,\dots,n+1 \,.
\end{equation}
where $g^{AB}$ is the inverse of the metric~(\ref{metric_hypersphere}). The Hamiltonian then reads
\begin{equation}
\begin{aligned}
\mathcal H =& \frac 1 2 \left[ p_0^2 - J^2 - \tan^{-2} \beta \left( L_1^2 + \sin^{-2} \theta^1 L_2^2 + \dots + \sin^{-2} \theta^1 \dots \sin^{-2} \theta^{n+1} L_{n+1}^2 \right) \right]  \\
& - v_0^{\frac 4 3}  \, e^{- \frac{2}{\sqrt{3}} x^0} \, C(\beta,\theta^{n+1}) + v_0^2 e^{-  \sqrt{3} x^0} \, V(\beta,\theta^1 ,\dots , \theta^n)  \,.
\end{aligned}
\end{equation}
The equations of motion are then
\begin{equation}
\begin{aligned}
& \dot x^0 = p_0 \,,&
& \dot p_0 = 
 - \frac{2}{\sqrt{3}} v_0^{\frac 4 3}  \, e^{- \frac{2}{\sqrt{3}} x^0} \, C(\beta,\theta^{n+1}) +\sqrt{3} v_0^2 e^{-  \sqrt{3} x^0} \, V(\beta,\theta^1 ,\dots , \theta^n) \,, 
\\
&
\dot \beta = - \cos^2\beta \, J \,,
&
&\dot J = -\cot^3 \beta  g^{ab} L_a L_b + v_0^{\frac 4 3}  \, e^{- \frac{2}{\sqrt{3}} x^0} \, \frac{\partial C(\beta,\theta^{n+1})}{\partial \beta} -  v_0^2 e^{-  \sqrt{3} x^0} \, \frac{\partial V(\beta,\theta^1 ,\dots , \theta^n)}{\partial \beta}\,,
\\
&\dot \theta^a = -g^{ab} L_b & & \dot L_a = + \frac 1 2 \frac{\partial g^{AB}}{\partial \theta^a} L_A L_B + \delta^a_{n+1} v_0^{\frac 4 3}  \, e^{- \frac{2}{\sqrt{3}} x^0} \, \frac{\partial C(\beta,\theta^{n+1})}{\partial \theta^{n+1}} -  v_0^2 e^{-  \sqrt{3} x^0} \, \frac{\partial V(\beta,\theta^1 ,\dots , \theta^n)}{\partial \theta^a} \,,  
\end{aligned}
\end{equation}
and, expressed in terms of $\beta$:
\begin{equation}
\begin{aligned}
&
\frac{ \d x^0}{\d \beta} = - \frac{p_0}{\cos^2 \beta J} \,,
\qquad 
\frac{ \d  \theta^a}{\d \beta} =  \frac{g^{ab} L_b}{\cos^2 \beta J}  \,,
\\
&
\frac{ \d p_0}{\d \beta} = 
 \frac{2}{\sqrt{3}} \frac{v_0^{\frac 4 3}  \, e^{- \frac{2}{\sqrt{3}} x^0} \, C(\beta,\theta^{n+1})}{\cos^2\beta J} - \sqrt{3} \frac{v_0^2 e^{-  \sqrt{3} x^0} \, V(\beta,\theta^1 ,\dots , \theta^n)}{\cos^2 \beta J}  \,,
\\
&\frac{ \d J}{\d \beta} = \frac{\cos \beta}{\sin^3 \beta J}  g^{ab} L_a L_b - \frac{v_0^{\frac 4 3}  \, e^{- \frac{2}{\sqrt{3}} x^0} \, \frac{\partial C(\beta,\theta^{n+1})}{\partial \beta}}{\cos^2 \beta J} +
\frac{  v_0^2 e^{-  \sqrt{3} x^0} \, \frac{\partial V(\beta,\theta^1 ,\dots , \theta^n)}{\partial \beta}}{\cos^2 \beta J}\,,
\\
&
 \frac{ \d  L_a}{\d \beta} = - \frac 1 2 \frac{\frac{\partial g^{AB}}{\partial \theta^a} L_A L_B}{\cos^2 \beta J} -
  \delta^a_{n+1} \frac{v_0^{\frac 4 3}  \, e^{- \frac{2}{\sqrt{3}} x^0} \, \frac{\partial C(\beta,\theta^{n+1})}{\partial \theta^{n+1}}}{\cos^2 \beta J} +
   \frac{ v_0^2 e^{-  \sqrt{3} x^0} \, \frac{\partial V(\beta,\theta^1 ,\dots , \theta^n)}{\partial \theta^a}}{\cos^2 \beta J} \,. 
\end{aligned}
\end{equation}
We can prove the well-posedness of the above equations at $\beta = \pi/2$ just like before. Moreover, during a `Kasner-like' epoch in which  $L_A^2  >>  - v_0^{\frac 4 3}  \, e^{- \frac{2}{\sqrt{3}} x^0} \, C(\beta,\theta^{n+1}) + v_0^2 e^{-  \sqrt{3} x^0} \, V(\beta,\theta^1 ,\dots , \theta^n)  $ for all $A$, the equations turn into equations for geodesics on the $(n+2)$-sphere expressed in terms of one angular coordinate.

\section{Outlook and conclusions}

In 1976 Steven Hawking wrote \cite{HawkingPRD1976}:
\begin{quote}
A singularity can be regarded as a place where
there is a breakdown of the classical concept of
space-time as a manifold with a pseudo-Reimannian
metric. Because all known laws of physics
are formulated on a classical space-time background,
they will all break down at a singularity.
\emph{This is a great crisis for physics because it means
that one cannot predict the future: One does not
know what will come out of a singularity.}
\end{quote}
This appparent loss of predictivity is one of the most disturbing features of General Relativity, and is often used to argue that the (classical) theory must be incomplete. It is intimately connected to far-reaching issues like black hole unitarity, and the dominant opinion nowadays is that one should look to quantum effects for a resolution of singularities and a restoration of predictivity.

In~\cite{ThroughTheBigBang} my coauthors and I challenged the idea that gravitational singularities mark a loss of predictivity of the classical theory by providing for the first time an example of a singularity where Einstein's equations satisfy an existence and uniqueness theorem. This means that each solution can be continued in a unique way past the singularity, and all the information that the solution is carrying is conserved through it.

The example we found in~\cite{ThroughTheBigBang} might be nothing more than a curiosity were it not for the fact that it describes a situation that is very close to the one that is realized our own universe: a homogeneous (but not necessarily isotropic) cosmology with a massless scalar field. This motivates further exploration of the validity of the continuation result with the goal of testing this conjecture: that it is possible to continue through the singularity at the beginning of our universe into another universe with an opposite time direction and spatial orientation with all the information about our universe conserved through the Big Bang. There is a long list of limitations of the original result that need to be relaxed in order to test the conjecture: for example, the homogeneity assumption, for which the BKL conjecture~\cite{BKL}  gives hope that relaxing it won't change the result. Moreover, the matter content should be enlarged to include Standard Model fields (gauge fields and fermions), and one should check whether all the information these fields carry is conserved through the Big Bang. Last, the scalar field should be modified in order to match the type of scalars we expect to be present in primordial cosmology. The prime candidate is an inflaton with a slow-roll inflationary potential.

The present paper represents the first step towards the generalization of the continuity result to realistic cosmology. The simplest generalization is to consider an inflationary potential for the scalar field. I'm pleased to report a first success of the conjecture: it holds true also when the scalar field has a nonzero potential provided the potential doesn't grow too fast when the scalar field goes to infinity. In particular, the conjecture holds true for the most emblematic choice of scalar field potential: Starobinsky's model, which has two particularly attractive features. First, it sits right in the middle of the current experimental cosmological constraints on inflationary models~\cite{Planck2018-X}. Second, the model follows from the inclusion of the first-order quantum corrections to Einstein's gravity and therefore it is a natural consequence of what we know of quantum gravity treated as an effective field theory. In the last section of the present paper I considered a further generalization to an arbitrary number of scalar fields with arbitrary potential. Again the result holds given that the potential does not grow too fast.

The next natural steps will be to test the conjecture with homogeneous fermions and gauge fields and with inhomogeneous perturbations of the metric. This will be the subject of future work.

\providecommand{\href}[2]{#2}\begingroup\raggedright\endgroup

\newpage

\appendix

\section{Appendix: Misner variables}
\label{Appendix_A}

In this Appendix I will explicitly show how to perform the reduction from the infinite-dimensional phase space of ADM variables $g_{ij}$, $p^{ij}$ to the finite-dimensional one of Misner variables $x$, $y$, $k_x$, $k_2$, which describe the degrees of freedom of a homogeneous, but not necessarily isotropic, universe with $S^3$ spatial topology. The starting point is the existence of three translation-invariant one-forms on $S^3$ and their associated dual vector fields:
\begin{equation}\label{Translation-invariant-one-forms-vector-fields-S3}
\left\{ \begin{aligned}
\sigma^1 &=  \sin r \, \d \theta - \cos r \, \sin \theta \, \d \varphi
\\
\sigma^2 &=  \cos r \, \d \theta + \sin r \, \sin \theta \, \d \varphi 
\\
\sigma^3 &=  - \d r   - \cos   \theta \, \d \varphi 
\end{aligned} \right.  \,,
\qquad
\left\{\begin{aligned}
\chi_1 &=  \cos r \, \cot \theta \, \partial_{ r} +\sin r \, \partial_{ \theta} - \cos r \, \csc \theta \, \partial_{  \phi} 
\\
\chi_2 &=  - \sin r \, \cot \theta \partial_{ r} +  \csc \theta  \left( \cos r \partial_{ \theta} + \sin r \, \partial_{ \phi } \right)
\\
\chi_3 &=  - \partial_{ r } 
\end{aligned}\right.\,,
\end{equation}
which are such that $\chi_b^i \sigma_j^a = \delta^a_b$. $ r \in (0,\pi)$, $\theta \in (0,\pi)$,  $\varphi \in (0,2\pi]$ are the standard spherical coordinates on $S^3$. Then any translation-invariant tensor can be written as linear combinations of tensor products of~(\ref{Translation-invariant-one-forms-vector-fields-S3}). In particular, we can write the 3-metric $g_{ij}$ (a covariant 2-tensor) and the conjugate momenta $p^{ij}$ (a contravariant 2-tensor \emph{density}) as
\begin{equation}
\begin{aligned}
&g_{ij} =  \sum_{a,b=1}^3 q_{ab} \, \sigma^a_i \,\sigma^b_j \,,
\qquad
p^{ij} = | \sigma^1 \wedge \sigma^2 \wedge \sigma^3 | \sum_{a,b=1}^3 p^{ab} \, \chi^i_a \,\chi^j_b \,,
\end{aligned}
\end{equation}
where the factor $| \sigma^1 \wedge \sigma^2 \wedge \sigma^3 |$ ensures that $p^{ij}$ transforms like a tensor density under diffeomorphisms. Symmetry of  $g_{ij}$ and $p^{ij}$ implies that  $q_{ab}$ and $p^{ab}$ are symmetric matrices too. We have reduced the infinite degees of freedom of $g_{ij}$ to the six independent components of a symmetric real matrix, and the same for $p^{ij}$.

Under this ansatz, we can explicitly perform the integrals in the symplectic potential~(\ref{ADM-symplectic-potential}):
\begin{equation}
\Theta = 4 \pi^2 \sum_{a=1}^3 p^{ab} \delta q_{ab} \,,
\end{equation}
and the ADM constraints~(\ref{ADM-constraints}):
\begin{equation}\label{Homogeneous-Hamiltonian-constraint}
\mathcal H[N] = \frac{4 \pi^2 N}{\sqrt{|\det q|}} \left[ {\sfrac 1 2} (\text{tr} q)^2  - \text{tr} (q^2)  -  p^{ab}q_{bc} p^{cd} q_{da} + {\sfrac 1 2} \left( p^{ab} q_{ab} \right)^2 - {\sfrac 1 2} \pi_\phi^2
- |\det q|\, V(\phi)\right] \,,
\end{equation}
\begin{equation}\label{Homogeneous-diffeo-constraint}
\mathcal D [N_i] = \int \d^3 x \mathcal H_i N^i = 4 \pi^2 N
\epsilon_{ab}{}^c \xi^a p^{bd} q_{dc} \,,
\end{equation}
where of course we implemented the homogeneous ansatz also for the lapse, $N = \text{\it const.}$, the shift vector, $N_i = \xi_a \sigma^a_i$ and the scalar field $\phi = \text{\it const.}$.
Varying the diffeomorphism constraint~(\ref{Homogeneous-diffeo-constraint}) with respect to the three components of $\xi^a$ leads to a particularly simple constraint:
\begin{equation}
[p,q]^a{}_b \approx 0 \,,
\end{equation}
which tells us that the matrices $q_{ab}$ and $p^{ab}$ commute. Since they are both symmetric, this implies that they are simultaneously diagonalizable. In fact, one can verify~\cite{FlavioSDbook} that imposing that $q_{ab}$ is diagonal (\emph{i.e.} $q_{12} = q_{23} = q_{13} =0$) is a good gauge-fixing for the constraint~(\ref{Homogeneous-diffeo-constraint}) and immediately implies that $p^{ab}$ is diagonal too ($p^{12} = p^{23} = p^{13} =0$).

It is now convenient, in order to make the kinetic term quadratic, to introduce the Ashtekar--Henderson--Sloan variables~\cite{AHS}:
\begin{equation}\label{AHSvariables}
\begin{aligned}
&p^{ab} = \text{diag} ( P_1 C_1 , P_2 C_2 , P_3 C_3 )  \,, &    & q_{ab} = \text{diag} ( C_1 , C_2 , C_3)   \,. & 
\end{aligned}
\end{equation}
Then the symplectic form and the Hamiltonian constraint~(\ref{Homogeneous-Hamiltonian-constraint}) take the form (after a trivial rescaling and introducing $P=\sum_a P_a$, $C = \sum_a C_a$)
\begin{equation}
\Theta = \sum_{a=1}^3\frac{ P_a  \,\delta C_a}{C_a} \,, \qquad 
{\sfrac 1 2} C^2 - \sum_{a=1}^3 C_a^2  +  {\sfrac 1 2} P^2 -
 \sum_{a=1}^3 (P_a)^2  - {\sfrac 1 2} \pi_\phi^2   -  C_1 C_2 C_3 \, V(\phi)
   \approx 0\,.
\end{equation}
We can make a coordinate transformation that simultaneously diagonalizes the kinetic energy and symplectic form and separates the scale degree of freedom from the remaining `shape' degrees of freedom. 
This is based on Jacobi coordinates for the $P_a$ variables 
and for the log of the $C_a$ variables (we can take their square roots and their logs because $C_a >0$ $\forall a$),
\begin{equation}
\begin{aligned}
& k_x = \frac{P_2 - P_1}{\sqrt 2}\,, &
& k_y = \sqrt{\frac 2 3}  \left(  \frac{P_2 + P_1}{2} - P_3  \right) \,, &
& D = \frac 1 3 \left( P_1 + P_2 + P_3 \right) \,, &
\\
& x = \frac{1}{\sqrt 2}  \log \left(\frac{C_2}{C_1} \right)\,, &
& y = \frac 2 3 \log  \left(  \frac{\sqrt{C_2 C_1}}{C_3}\right) \,, &
& v =   \sqrt{ | C_1 C_2 C_3 |}    \,. &
\end{aligned}
\end{equation}
Notice that the $a$ and $b$ variables are scale-invariant while $v$ is not ($v$ is proportional to the volume of the spatial slice).
The inverses of the above relations are
\begin{equation}\label{InvertedAHSvariables}
\begin{aligned}
& P_1= \frac{k_y}{\sqrt{6}} - \frac{k_x}{\sqrt{2}} + D \,, &
&  P_2= \frac{k_x}{\sqrt{2}} + \frac{k_y}{\sqrt{6}}+ D \,, &
&  P_3= -\sqrt{\frac{2}{3}} k_y + D  \,, &\\
& C_1= e^{- x/\sqrt{2} + y/\sqrt{6} }  v^{\frac 2 3} \,, &
&  C_2= e^{ x/\sqrt{2} + y/\sqrt{6} }  v^{\frac 2 3} \,, &
&  C_3= e^{- \sqrt{\frac{2}{3}} y } v^{\frac 2 3} \,. &
\end{aligned}
\end{equation}
Replacing the last identities into the symplectic potential we get
\begin{equation}
\Theta =  k_x \, \delta x + k_y \, \delta y + 2 \frac{D}{v} \delta v \,,
\end{equation}
and the Hamiltonian constraint now takes the form
\begin{equation}
\mathcal H =
{\sfrac 3 2} D^2 - k_x^2 - k_y^2
- v^{4/3} \, U(x,y)  - v^2 \, V(\phi)
   \approx 0\,,
\end{equation}
where
\begin{equation}
U(x,y)= f(-2y)+f(\sqrt 3 x +y)+f(-\sqrt 3 x +y) \,,
\qquad f(z) =  {\sfrac 1 2} e^{- \frac {2 z} {\sqrt 6}} - e^{- \frac z {\sqrt 6}}\,.
\end{equation}
As a last step, it is customary to introduce the canonically conjugate variable to the volume, the \emph{York time} $\tau = 2 D/v$. Then the symplectic potential and the Hamiltonian constraint take the form
\begin{equation}
\Theta =  k_x \, \delta y + k_y \, \delta y + \tau \delta v \,,
\qquad
\mathcal H =
{\sfrac 3 8} \tau^2 \, v^2  - k_x^2 - k_y^2
- v^{4/3} \, U(x,y)  - v^2 \, V(\phi)
   \approx 0\,.
\end{equation}

\end{document}